\documentclass{sig-alternate}
\usepackage{amssymb}
\usepackage{amsmath}
\usepackage{amsfonts}
\usepackage{graphicx}
\usepackage{subfigure}
\usepackage{algorithm}
\usepackage{algorithmic}
\usepackage{url}
\usepackage{multirow}
\usepackage{mdwlist}
\usepackage{enumitem}
\usepackage[square,sort,comma,numbers]{natbib}

\begin{document}

\title{Friendship Prediction in Composite Social Networks}

\numberofauthors{1}

\author{\alignauthor
{ Erheng Zhong$^\dag$, Evan Wei Xiang$^*$, Wei Fan$^\ddag$, Nathan Nan Liu$^\#$, and Qiang Yang$^\dag$$^\ddag$}\\
\affaddr {\normalsize $^\dag$Hong Kong University of Science and Technology, Hong Kong}\\
\affaddr {\normalsize $^\ddag$Huawei Noah's Ark Lab, Hong Kong}\\
\affaddr {\normalsize $^*$Baidu Inc., Shenzhen, China}\\
\affaddr {\normalsize $^\#$Yahoo! Labs, California, USA}\\
\email{{\normalsize \{ezhong,qyang\}@cse.ust.hk, xiangwei@baidu.com, evandavid.fanwei@huawei.com, nanliu@yahoo-inc.com}}
}

\date{}

\maketitle

\newdef{theorem}{Theorem}
\newdef{definition}{Definition}
\newdef{lemma}{Lemma}
\newdef{assumption}{Assumption}

\begin{abstract}
Friendship prediction is an important task in social network analysis (SNA). It can help users identify friends and improve their level of activity. Most previous approaches predict users' friendship based on their historical records, such as their existing friendship, social interactions, etc. However, in reality, most users have limited friends in a single network, and the data can be very sparse. The sparsity problem causes existing methods to overfit the rare observations and suffer from serious performance degradation. This is particularly true when a new social network just starts to form. We observe that many of today's social networks are ``composite'' in nature, where people are often engaged in multiple networks. In addition, users' friendships are always correlated, for example, they are both friends on Facebook and Google+. Thus, by considering those overlapping users as the bridge, the friendship knowledge in other networks can help predict their friendships in the current network. This can be achieved by exploiting the knowledge in different networks in a collective manner. However, as each individual network has its own properties that can be incompatible and inconsistent with other networks, the naive merging of all networks into a single one may not work well. The proposed solution is to extract the common behaviors between different networks via a hierarchical Bayesian model. It captures the common knowledge across networks, while avoiding negative impacts due to network differences. Empirical studies demonstrate that the proposed approach improves the mean average precision of friendship prediction over state-of-the-art baselines on nine real-world social networking datasets significantly.
\end{abstract}

\section{Introduction}
With the recent development of Web 2.0, online social networks, such as Facebook, Google+, QQ and Tencent's Microblog in China, are becoming increasingly popular. Friendship prediction, which aims to predict the relationship between two users, plays an important role in analyzing these networks and promoting the level of user activity, i.e., encouraging people to interact with their friends. Besides inferring relationship~\cite{Wang:2011:HMS:2020408.2020581}, friendship prediction also helps predict profiles~\cite{Mislove:2010:YYK:1718487.1718519}, detect users' influence~\cite{DBLP:conf/icwsm/ChaHBG10} and perform behavior targeted advertising~\cite{Liu:2011:LBT:2063576.2063838}, etc. One important observation is that, online social networks are usually \emph{composite}~\cite{Zhong:2013:MDC:2487575.2487652}, where different social networks overlap with each other. People may engage in many different networks simultaneously for different purposes. For example, people usually use Facebook and Google+ to communicate with their friends, but interact with others who share similar video interests on Youtube.

Previous research works based on single-network friendship prediction may fail to predict correctly due to the data sparsity problem. Each user may have only a few neighbors and this is particularly true for a new network that is starting to form. Thus, each single network may capture only partial aspects of users' social interests. Learning from such sparse network can cause the model to overfit rare observed links. For example, neighborhood models may suggest the friends of friends to a given user. But if the user builds links with only a few other users, most of his/her friendships that rely on unobserved friends cannot be discovered. In addition, information in a single network can be incomplete, but in the same time, different social networks may be correlated with each other. To be exact, the friendship generation process in the current network can be influenced by activities captured by some other networks. For example, if two people become friends in Google+ for similar hobbies on videos, they are likely follow each other on Youtube as well. Without priors from other networks, such links may not be predicted due to the incomplete knowledge in any individual network.

However, the composite property, that users and links across different networks overlap, sheds light on solving the above problems. By using the unified identity, such as Gmail account for Google+ and Youtube, and QQ number for QQ and Tencent's Microblog, common users can be identified across networks within a company. In addition, several works have been proposed to link users across different institutes based on their profiles~\cite{Liu:2013:WNU:2433396.2433457,Yuan:2013:MSRA,Liu:2013:CM}. Consequently, by considering the common users in different networks as the bridge, knowledge in other networks can be transferred to the current network to overcome the ``sparsity'' challenge. For example, the auxiliary knowledge from other networks can be exploited in auxiliary priors, to model the generating process in the current network. Nevertheless, one cannot simply merge multiple networks due to network differences. First, different networks have different properties, such as different levels of density, degree distributions, clustering coefficient or diameters. If we merge two networks together, their specific network structures can be destroyed. For example, if we merge a dense network and a sparse network directly, the knowledge in the sparse network will be hidden. Second, users play different roles and generate different communities in different networks. Two users holding a link in one network are not necessary neighbors in another network. Specifically, for a given users, his/her neighbors in different single networks can overlap but may not necessary be the same. For example, one person may link with some users on both Youtube and Google+. At the same time, he/she may link with some others on Youtube only if they share similar video interests, but will not become friends with them on Google+ as they are not familiar with each other.

Thus, the motivation of the proposed work is to model the current network with other networks together in order to benefit from the enriched knowledge, while resolving the network differences. The challenge here is how to distinguish the shared and specific knowledge across networks, given the sparse data in each single network. We propose a hierarchical mixed membership model, Composite Friendship Prediction (ComFP), to solve the problem. The proposed model integrates friendship knowledge over multiple different social networks collectively to help the prediction in the current network. First, ComFP utilizes an adaptive prior for each user to represent different users' global interests over all individual networks. Since this prior is related to all nested networks and thus it can encode the common knowledge across multiple networks. Second, ComFP introduces another network-specific prior to encode particular knowledge in the current network, in order to model the network differences. This prior helps avoid the knowledge in a sparser network to be overwhelmed by other denser networks. Finally, these two priors are combined together as a hybrid prior to build mixed membership models in individual networks. Generally, the knowledge in other networks guides the current model building through users' global interests, while the network-specific priors avoid negative impacts of network differences and simultaneously allow the current network to maintain its own properties. We propose a Gibbs sampler to construct mixed membership models and a Metropolis-Hastings sampler to infer the hybrid priors.

{\bf Problem Formulation} We define the problem of friendship prediction across multiple social networks in this section. The notations are summarized in Table~\ref{tab:Notation}. Let $\mathbb{G}=\{G_i=(U_i,E_i)\}_{i=1}^{N}$ denote a set of nested individual networks, where $G_i$ is the $i$-th individual network, $U_i$ is the user set of $G_i$, $E_i$ is the user relationship of $U_i$ and $N$ is the total number of individual networks. In addition, we define the whole user set as $\mathbf{U}$ and the whole link set as $\mathbf{E}$, where $\mathbf{U}=\cup\{U_i\}_{i=1}^N$ and $\mathbf{E}=\cup\{E_i\}_{i=1}^N$. The number of users in $\mathbf{U}$ is $n$ and the number of links in $E_i$ is $m_i$. An important property is that users in different individual networks overlap, and formally,
\begin{eqnarray}
U_i\cap{U_j}\neq\emptyset,\; \forall i\neq{j}, \; U_i\subseteq{\mathbf{U}}\; U_j\subseteq{\mathbf{U}}
\end{eqnarray}
This properties is easy to satisfy. For example, users of Google+ and Youtube are overlapping but not identical, e.g., two users may be friends on Google+ as well as followers on Youtube to each other. We notice that, on one hand, users' behaviors in one network can be influenced by their status or activities in other networks. One the other hand, different social networks have different properties, such as diameters, clustering coefficients, etc. The objective in this paper is to predict how likely an unobserved edge $e^k_{ij} \notin E_k$ exists between an arbitrary user pair $(u_i,u_j)$ in all individual networks using the data from $\mathbb{G}$, where it can benefit from the enriched information from multiple networks while avoiding negative impacts resulting from network differences.

    \begin{table}[t]
    \vskip -0.05in
    \caption{Definition of Notations}
    \begin{center}
    \begin{scriptsize}
    \begin{tabular}{|l|l|}
    \hline
    Notation   & Notation Description                 \\
    \hline
    \multicolumn{2}{|c|}{Data} \\
    \hline
    $\mathbf{U}=\cup\{U_i\}_{i=1}^N$  & User Set \\
    $\mathbf{E}=\cup\{E_i\}_{i=1}^N$  & Link Set \\
    $\mathbb{G}=\{\mathbf{U},\mathbf{E}\}$    & Nested Networks, $\{G_i\}_{i=1}^N$              \\
    $N$             & Number of individual networks                                             \\
    $n$             & Number of users            \\
    $m_i$           & Number of links in the $i$-th network \\
    \hline
    \multicolumn{2}{|c|}{Model} \\
    \hline
    $Dir(\cdot)$    & Dirichlet distribution          \\
    $Mult(\cdot)$   & Multinomial distribution       \\
    $Bern(\cdot)$   & Bernoulli distribution          \\
    $Beta(\cdot)$   & Beta distribution          \\
    $\pi_{i}$       & Membership vector of user $u_i$ \\
    $\mathbf{z}_{i\rightarrow{j}}$ & Indicator vector form user $u_i$ to $u_j$    \\
    $\mathbf{B}$    & Community compatibility matrix      \\
    $K_d$           & Number of communities of $G_d$    \\
    $T$             & Number of latent features         \\
    $\alpha, \sigma$             & Priors         \\
    $\lambda$                    & Network-specific factors         \\
    $\mathbf{x}$                 & Users' latent factors         \\
    \hline
    \end{tabular}
    \label{tab:Notation}
    \end{scriptsize}
    \end{center}
    \vskip -0.25in
    \end{table}

\begin{figure*}[t]
\begin{small}
\centering \mbox{
\subfigure[Tencent-QQ]{\scalebox{0.4}{\includegraphics[width=\columnwidth]{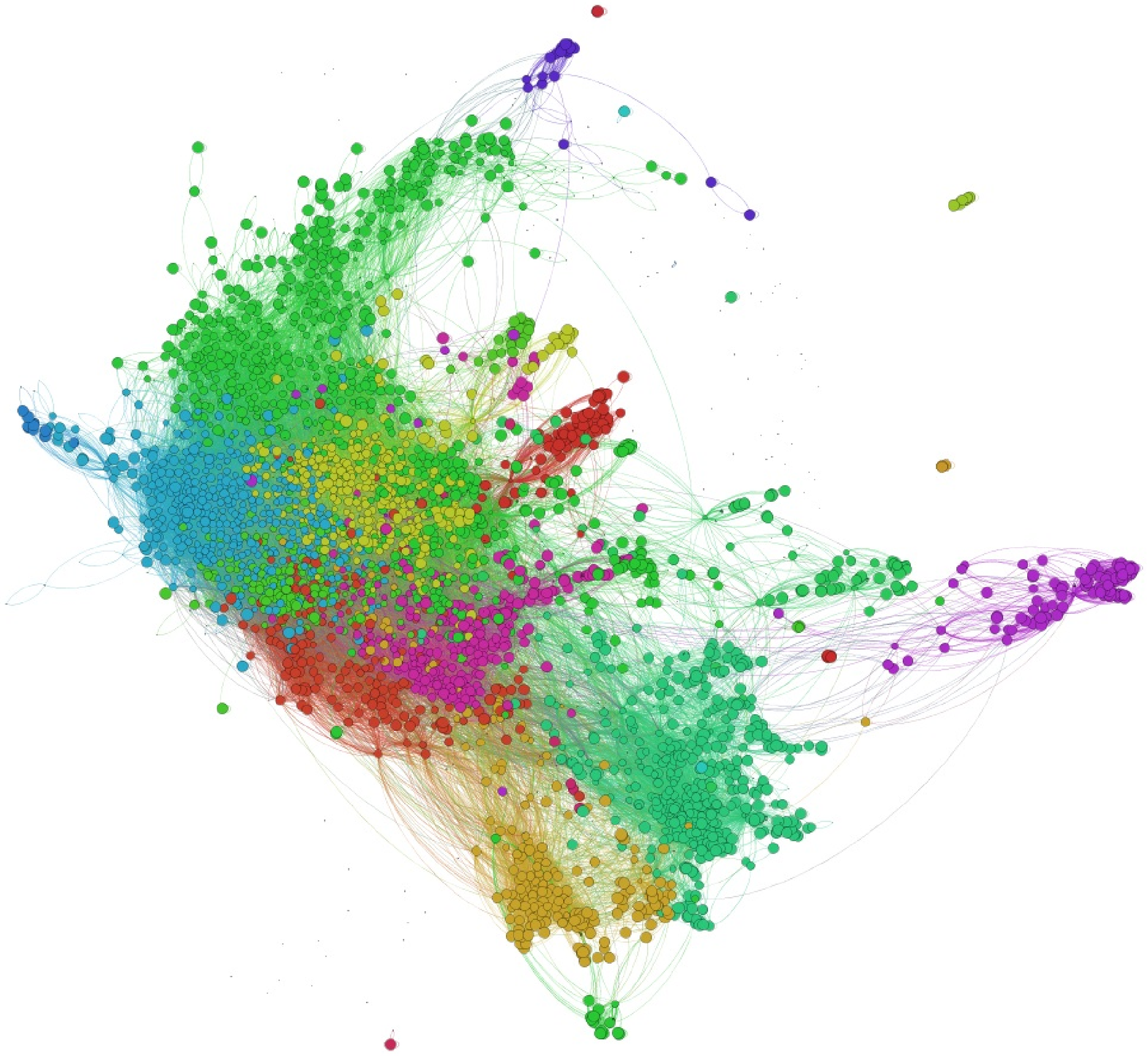}}}
\subfigure[Tencent-Microblog]{\scalebox{0.4}{\includegraphics[width=\columnwidth]{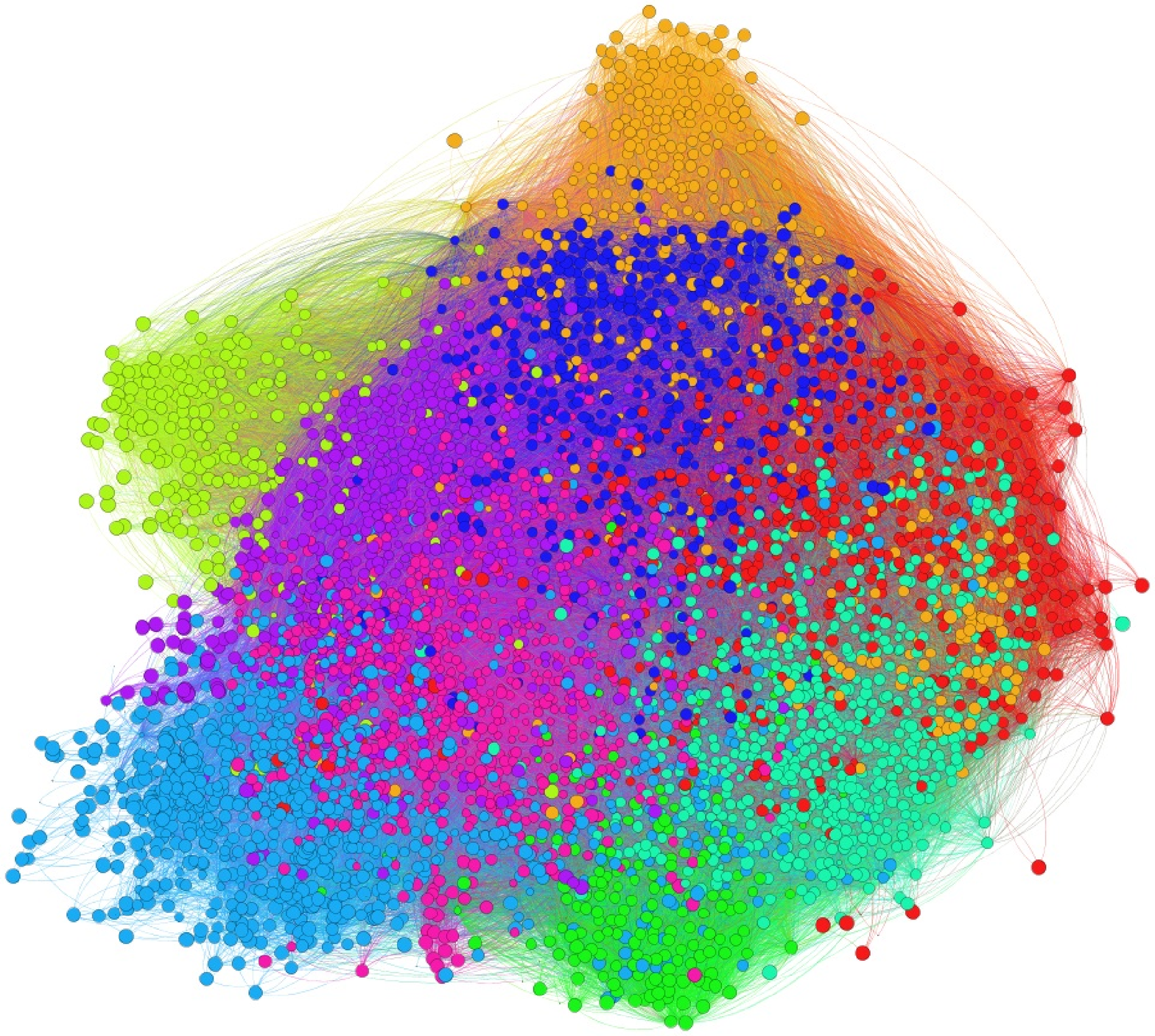}}}
\subfigure[Douban-Online]{\scalebox{0.4}{\includegraphics[width=\columnwidth]{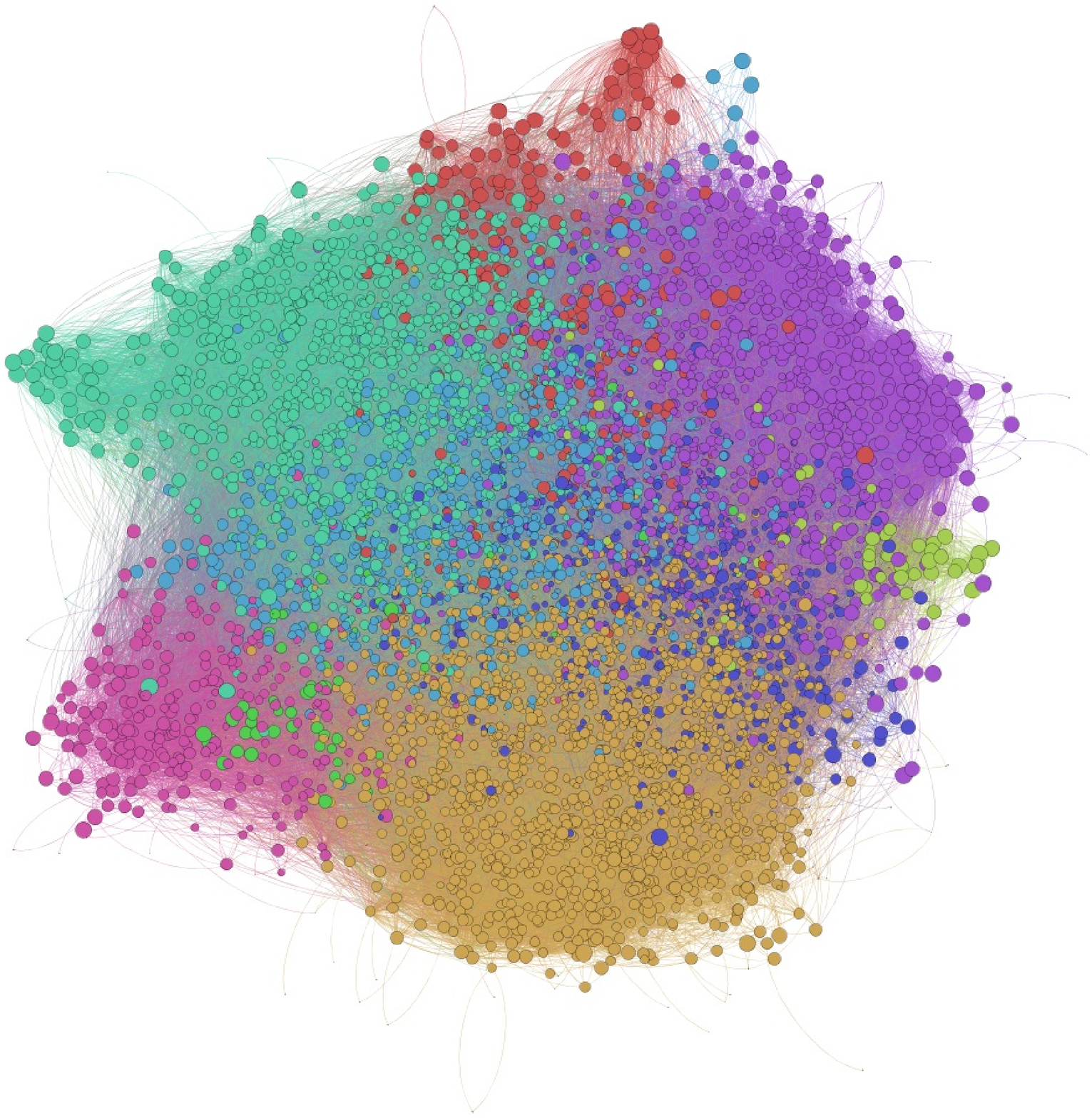}}}
\subfigure[Douban-Offline]{\scalebox{0.4}{\includegraphics[width=\columnwidth]{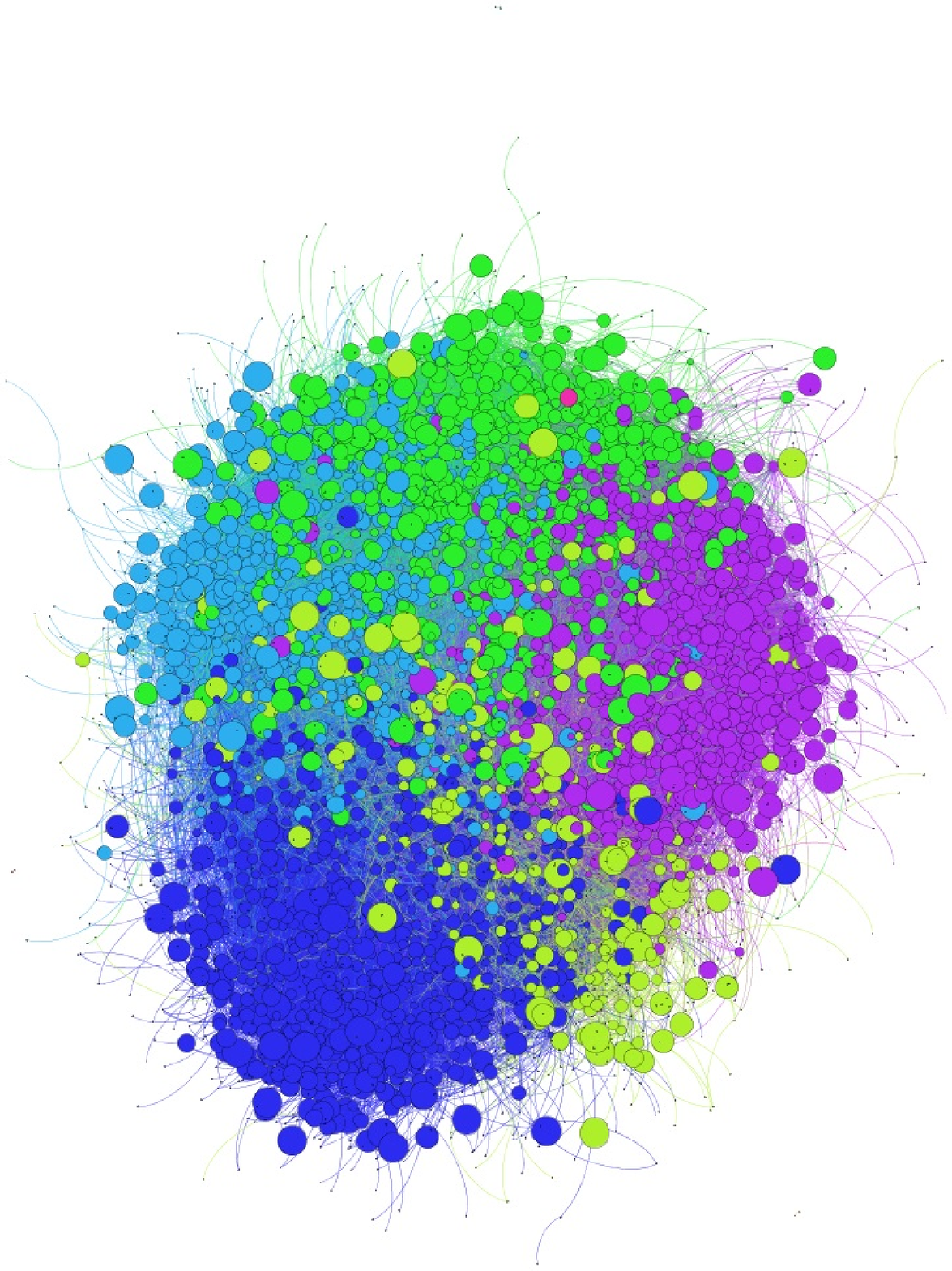}}}
}
\vskip -0.15in
\caption{\scriptsize Network Visualization. Different colors represent different communities. The size of each node presents each user's membership of the community, e.g., larger size indicates higher probability of the user belonging to this community. The communities are detected by the modularity maximization~\cite{Leskovec:2008:SPC:1367497.1367591}.} \label{fig:diff}
\vskip -0.15in
\end{small}
\end{figure*}

\section{Friendship Prediction across Multiple Social Networks}
We propose a novel mixed membership model, Composite Friendship Prediction (ComFP), to predict the friendships across multiple social networks collectively. In the following, we firstly describe the foundation of ComFP, Mixed Membership Stochastic Blockmodels (MMSB)~\cite{Airoldi:2008:MMS:1390681.1442798}, and then present the details of the composite modeling. Finally, we propose an efficient Gibbs sampling method to infer the latent variables. To help understand the motivation of the proposed model, we visualize two sub networks from Tencent and Douban in Figure~\ref{fig:diff}. The Tencent's networks contain users who are 2-hops away from the first author of the paper, as well as their relationships in Tencent's instant messaging network QQ and Microblog network. The Douban's networks contain a random subset of users from the crawled dataset and their online and offline relationships. Obviously, the community structures and users' memberships to communities in different networks can be quite different.

\subsection{Background of MMSB}
MMSB assumes that each user $u_i\in\mathbf{U}$ has a latent mixture of $K$ roles, which determine the membership of $K$ communities in the network $\mathcal{G}$. We denote this mixture as a normalized $K\times{1}$ vector $\pi_i$, which formalizes the notion of node multi-functionality. In MMSB, these vectors are drawn from some priors $p(\pi)$, such as Dirichlet distribution~\cite{Airoldi:2008:MMS:1390681.1442798} and Logistic-Normal distribution~\cite{Fu:2009:DMM:1553374.1553416}. In addition, MMSB generates a $K\times{K}$ community relation matrix $\mathbf{B}$ from some priors, like Beta distribution $Beta(\gamma_0,\gamma_1)$. $\mathbf{B}$ represents the probability of having a connection from a user in a community to another user in another community. Given the vector $\pi_i$ of each user $u_i$, the network edge $e_{ij}$ is generated stochastically as follows:
\begin{itemize}[noitemsep,topsep=0pt,parsep=0pt,partopsep=0pt]
  \item For each pair of users $(u_i,u_j)\in\mathbf{E}$ in the network $\mathcal{G}$:
  \begin{itemize}
    \item Draw indicator for $u_i$, $\mathbf{z}_{i{\rightarrow}j}\sim{Mult(\pi_{i})}$
    \item Draw indicator for $u_j$, $\mathbf{z}_{j{\rightarrow}i}\sim{Mult(\pi_{j})}$
    \item Sample the link, $e_{ij}\sim{Bern(\mathbf{z}_{i{\rightarrow}j}^T\mathbf{B}\mathbf{z}_{j{\rightarrow}i})}$
  \end{itemize}
\end{itemize}
where $\mathbf{z}_{i\rightarrow{j}}$ and $\mathbf{z}_{j\rightarrow{i}}$ are two $K\times{1}$ unit indicator vectors for the sender $u_i$ and the receiver $u_j$ respectively. Intuitively, the bilinear form $\mathbf{z}_{i\rightarrow{j}}^T\mathbf{B}\mathbf{z}_{j\rightarrow{i}}$ selects a single element of $\mathbf{B}$ as the probability of $e_{ij}$.

\begin{figure}[t]
\centering \mbox{
\subfigure[MMSB]{\scalebox{0.5}{\includegraphics[width=\columnwidth]{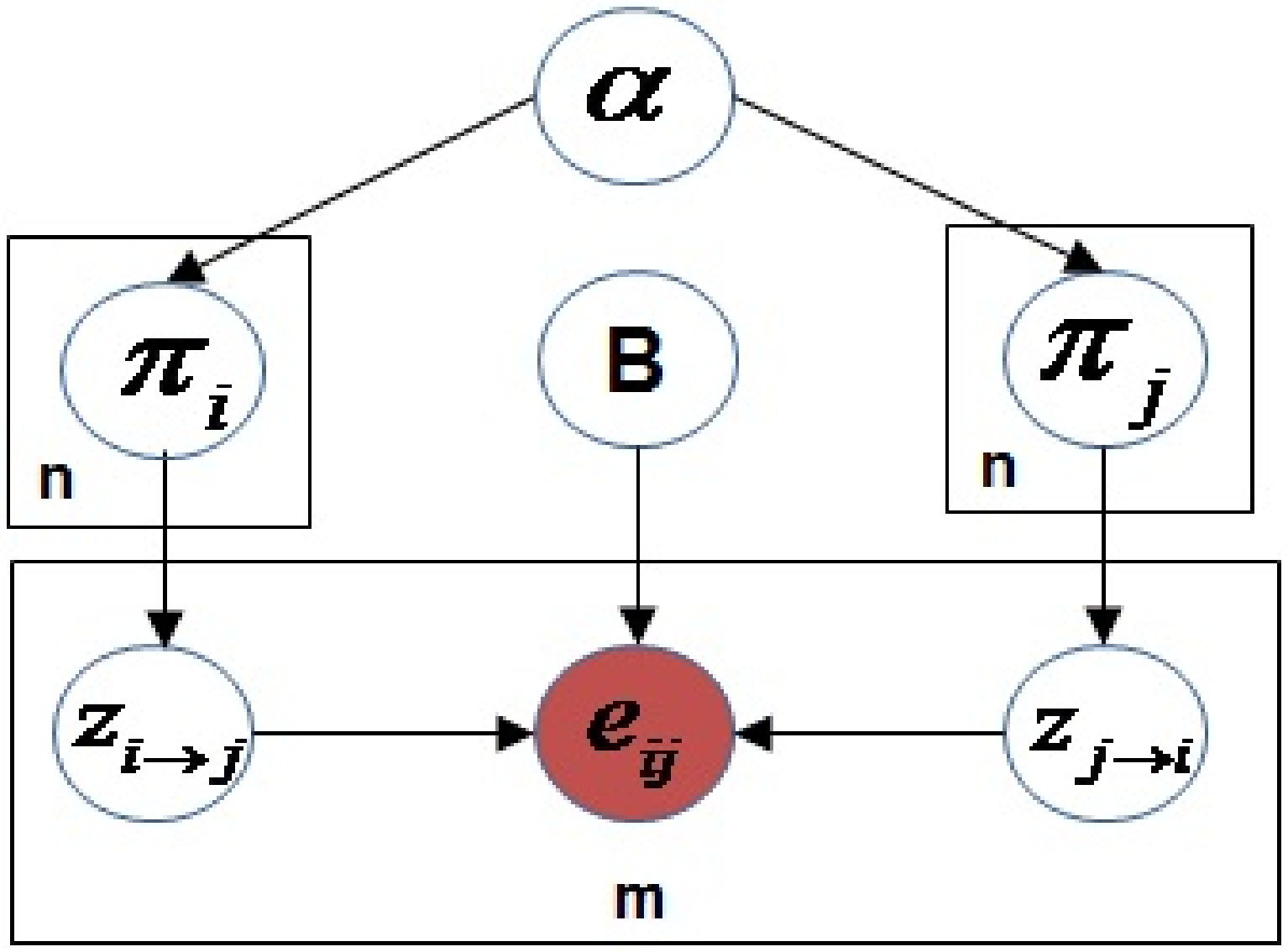}}}
\subfigure[ComFP]{\scalebox{0.5}{\includegraphics[width=\columnwidth]{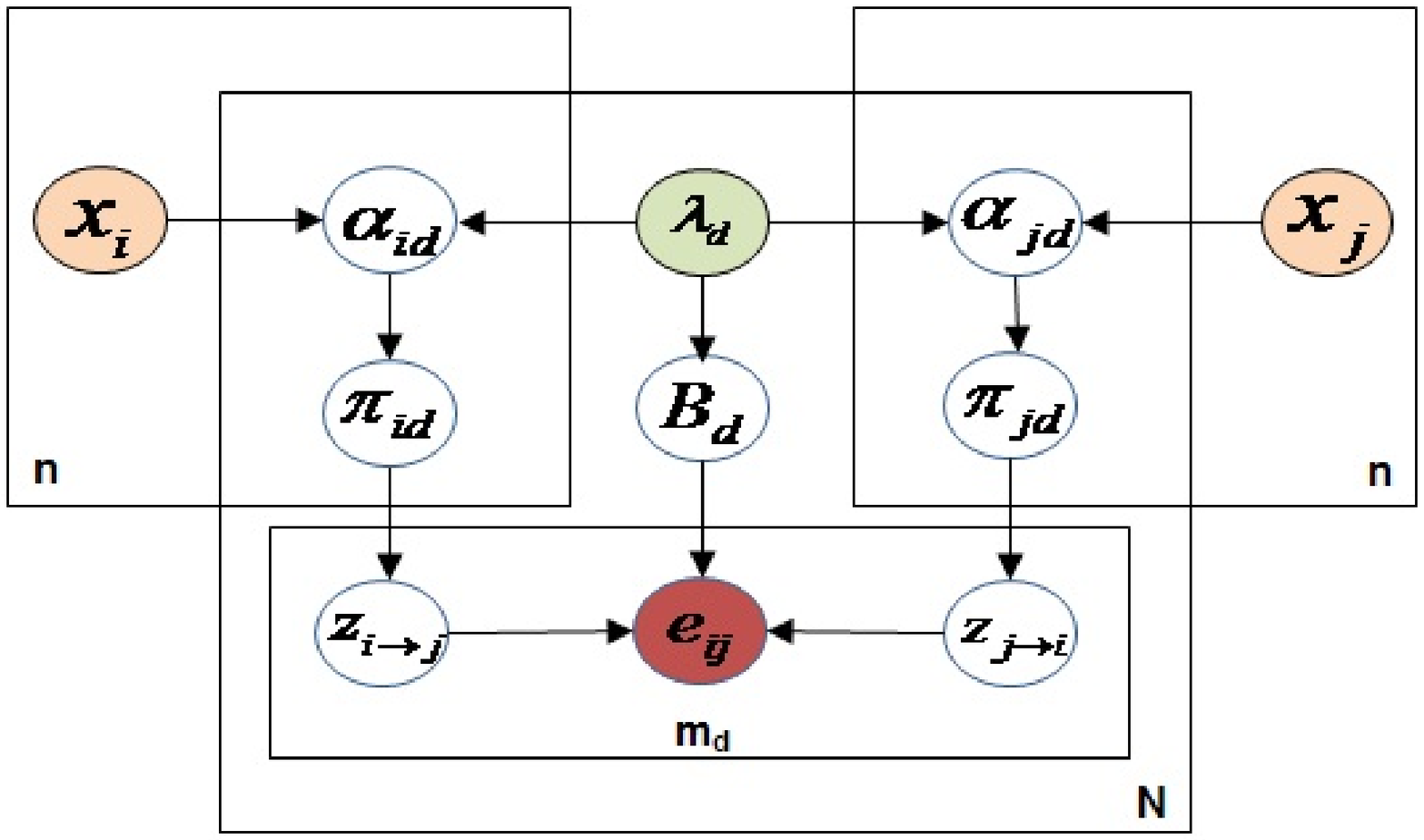}}}
}
\vskip -0.15in
\caption{\scriptsize Graphical Representations of MMSB and ComFP. $e_{ij}$ is the link between $u_i$ and $u_j$, $\mathbf{z}$ represents the community assignment, $\pi$ denotes each user's community membership, $\mathbf{B}$ is a community compatibility matrix, $\lambda_d$ is the feature-community mapping matrix in the network $G_d$, $\mathbf{x}_i$ denotes the latent features of user $u_i$ and $\alpha$ refers to the priors. In addition, $n$ is the number of users, $m$ refers to the number of link in an individual network and $N$ is the number of networks. Generally, the link between two users is determined by their community memberships as well as the community compatibility. } \label{fig:model}
\vskip -0.1in
\end{figure}

\subsection{Composite Friendship Prediction}
The motivation to model multiple nested social networks is to regularize the model of each individual network with the knowledge from other networks while taking care of network differences. Intuitively, we can divide the model into two parts: one part is to model the common knowledge which can transfer knowledge among different networks, and the other is to model network-specific knowledge. Here we implement the idea based on Mixed Membership Stochastic Block (MMSB) and propose a model called Composite Friendship Prediction (ComFP). The implementation is to introduce a hybrid prior $\alpha_{id}$ for each user $u_i$ in each network $G_d$. Specifically, the prior contains two components: $\lambda=\{\lambda_d\}_{d=1}^N$ and $\mathbf{x}=\{\mathbf{x}_i\}_{i=1}^{n}$, where $\mathbf{x}_i\in{\mathcal{R}^{1\times{T}}}$ represents the global interests of user $u_i$ and $\lambda_d\in{\mathcal{R}^{K_d\times{T}}}$ represents the network-specific characteristics of $G_d$.

Firstly, $\mathbf{x}_i$ can be understood as users' latent features, which represent users' social interests. These latent features reflect users' social behaviors, e.g., generating communities with other users. In addition, as shown in Figure~\ref{fig:diff}, the number of communities in each network $K_d$ can vary across networks and community structures in different networks can be different. That implies, in different single networks, different latent interests play different roles to generate communities. Thus, we introduce $\lambda_d$ here to map users' global latent features into network-dependent community membership. As an example, each user on Google+ and Youtube can be described based on different factors, where some factors represent users' interests on daily life and others indicate users' video interests. In ComFP, these factors are encoded in $\mathbf{x}_i$. Then, in Google+, the daily-life interests may play a more important role to generate users' links, while video interests may be more important in Youtube. This phenomenon is captured by $\lambda_d$ in ComFP. Consequently, after utilizing these two priors together, $\alpha_{id}=t(\mathbf{x}_i\lambda_d^T)$, where $t(x)=\log(1+e^{x})$, we can obtain a hybrid prior that indicates users' specific membership over communities in different individual networks. Importantly, this prior encodes the user-dependent and network-specific knowledge, which can exploit the auxiliary knowledge from other networks while distinguishing the network differences.

After that, this hybrid prior generates a vector $\pi_{id}$ for each user $u_i$ in each network $G_d$ following a Dirichlet distribution, in order to represent the user's membership over communities. At the same time, for each network, the network-specific prior $\lambda_d$ is exploited to generate the community compatibility matrix $\mathbf{B}_d$ for each network, which captures the relationship between each community, from a Beta distribution. The motivation is that, if two communities have similar feature mappings, users may have high probability to connect with each other. Then, for each user pairs, ComFP draws two asymmetric community indicator vectors according to the membership vectors for two users following a multinomial distribution. Consequently, a variable $\mathbf{z}_{i\rightarrow{j}}^T\mathbf{B}_d\mathbf{z}_{j\rightarrow{i}}$ is computed according to the indicator vectors and the community compatibility matrix. The variable indicates whether the two users $u_i$ and $u_j$ are linked or not. Finally, it samples the link $e_{ij}$ between two users $u_i$ and $u_j$ from a Bernoulli distribution based on this variable. Formally, the generative process is as follows,

\begin{itemize}[noitemsep,topsep=0pt,parsep=0pt,partopsep=0pt]
  \item For each network $G_d$:
  \begin{itemize}
    \item Draw a $K_d\times{T}$ feature matrix $\lambda_d\sim{\mathcal{N}(\mathbf{0},\sigma_d^2\mathbf{I})}$
    \item Draw $\mathbf{B}_d$, $\mathbf{B}^{ij}_d\sim{Beta(t(\lambda_d\lambda_d^T)^{ij},1)}$
  \end{itemize}
  \item For each user $u_i\in\mathbf{U}$:
  \begin{itemize}
    \item Draw a $1\times{T}$ latent feature vector $\mathbf{x}_i\sim{\mathcal{N}(\mathbf{0},\sigma_u^2I)}$
    \item For each network $G_d$:
    \begin{itemize}
      \item Generate a network-user prior $\alpha_{id}=t(\mathbf{x}_i\lambda_d^T)$
      \item Draw a membership vector $\pi_{id}\sim{Dir(\alpha_{id})}$
    \end{itemize}
  \end{itemize}
  \item For each user pair $(u_i,u_j)\in{E_d}$ in each network $G_d$:
  \begin{itemize}
    \item Draw indicator for $u_i$, $\mathbf{z}_{i\rightarrow{j}}\sim{Mult(\pi_{id})}$
    \item Draw indicator for $u_j$, $\mathbf{z}_{j\rightarrow{i}}\sim{Mult(\pi_{jd})}$
    \item Sample the link, $e^d_{ij}\sim{Bern(\mathbf{z}_{i\rightarrow{j}}^T\mathbf{B}_d\mathbf{z}_{j\rightarrow{i}})}$
  \end{itemize}
\end{itemize}

The graphical representations of MMSB and ComFP are shown in Figure~\ref{fig:model}. Obviously, MMSB does not incorporate the knowledge across networks. It may fail to infer users' community membership $\pi_i$ due to the sparse data, e.g., only a few links $e_{ij}$ for the user $u_i$ in the current network. An intuitive solution is to merge multiple networks simply and then apply MMSB. However, network differences, such as different community structures as shown in Figure~\ref{fig:diff}, are ignored in model construction, as the uniform prior is applied. The knowledge in a sparser network can be hidden and hence the community membership of each user cannot reflect the specific properties in this network. That implies a single $\pi$ cannot capture users' community memberships accurately in all single networks. Instead, ComFP captures the network-specific and user-dependent knowledge to model common knowledge and allows different users to own different community memberships in different networks. Specifically, common knowledge from other networks is embedded in each user's global prior $\mathbf{x}_i$ and another prior $\lambda_d$ adjusts users' community memberships and community-community relations $\mathbf{B}_d$ in different networks.

\subsection{Inference}
The inference process has two scenarios: one is to construct MMSB using Gibbs sampling and the other is to learn the hierarchical priors using a Metropolis-Hastings sampler. As joint inference is intractable due to the non-conjugate between hierarchical priors and the latent factors in MMSB, we perform alternate inference by calling Gibbs sampling and Metropolis-Hastings sampler in turns iteratively.

{\bf MMSB} We notice that although each MMSB in each single network is similar to the one described in~\cite{Airoldi:2008:MMS:1390681.1442798}, we assign a prior on $\mathbf{B}$ to restrict the probability of generating links between two communities, while the previous work does not. In addition, the original paper solved the MMSB using Variational EM, which may not approximate the true posterior distribution and thus we propose a Gibbs sampling method instead. First, let $\rho_{d}=t(\lambda_d\lambda_d^T)$, and then the pairwise community assignments in the $d$-th network can be written as
\begin{eqnarray}
&&p(\mathbf{z}^d|\alpha_d,\rho_d,E_d) \propto p(E_d|\mathbf{z}^d,\rho_d)p(\mathbf{z}^d|\mathbf{\alpha}_d) \\
&=&\int{p(E_d|\mathbf{z}^d,\mathbf{B}_d)dp(\mathbf{B}_d|\rho_{d})}\int{p(\mathbf{z}^d|\pi_d)dp(\pi_d|\mathbf{\alpha}_d)} \nonumber\\
&=&\prod_{k,k'}\frac{B(\rho_{d,k,k'}+\mathbf{n}_{k,k'})}{B(\rho_{d,k,k'})}\prod_{u_i}\frac{B(\alpha_{id}+\mathbf{n}_{i,\cdot})}{B(\alpha_{id})}\prod_{u_j}\frac{B(\alpha_{jd}+\mathbf{n}_{j,\cdot})}{B(\alpha_{jd})}\nonumber
\end{eqnarray}
where $B(\omega)=\frac{\prod_k\Gamma(\omega_k)}{\Gamma(\sum_k\omega_k)}$, $\mathbf{n}_{i,\cdot}$ denotes the total number of communities assigned to user $u_i$ and $\mathbf{\alpha}_d$ is a matrix including each user' prior in the $d$-th network. Then, after transformation and elimination, the pairwise posterior of the community distribution on the $d$-th network can be defined as
\begin{eqnarray}
\label{eq:gibbs}
&p(\mathbf{z}^d_{i\rightarrow{j}}=k, \mathbf{z}^d_{j\rightarrow{i}}=k'|\mathbf{\alpha}_d,\rho_{d,k,k'},\mathbf{z}^d_{\neg(i,j)}, E_d)\\
\propto&\prod_{k,k'}\Gamma(n^d_{i,k}+\alpha^k_{id})\Gamma(n^d_{j,k'}+\alpha^{k'}_{jd})\frac{B(\mathbf{n}^d_{k,k'}+\rho_{d,k,k'}+1)}{B(\rho_{d,k,k'}+1)}\nonumber \\
\propto&\prod_{k,k'}\Gamma(n^d_{i,k}+\alpha^k_{id})\Gamma(n^d_{j,k'}+\alpha^{k'}_{jd})\frac{\Gamma(n^d_{k,k',y_{(i,j)}}+\rho_{d,k,k'}+1)}{\Gamma(\sum_{y}(\rho_{d,k,k'}+1+n^d_{k,k',y}))} \nonumber\\
\propto&(n_{i,k}^{d,\neg{(i,j)}}+\alpha^{k}_{id})(n_{j,k'}^{d,\neg{(i,j)}}+\alpha^{k'}_{jd})\frac{n_{k,k',y_{(i,j)}}^{d,\neg{(i,j)}}+\rho_{d,k,k'}+1}{\sum_{y}(\rho_{d,k,k'}+1+n_{d,k,k',y}^{\neg{(i,j)}})}\nonumber
\end{eqnarray}
where $\mathbf{z}^d_{\neg(i,j)}$ denotes the set of community assignments without two assignments over the link between $u_i$ and $u_j$, $n^d_{i,k}$ represents the number of user $u_i$ picking community $k$ in the $d$-th network, and $n^d_{k,k',y}$ represents the total number of links in type $y$ with $(k,k')$ as the participating communities in the $d$-th network. In addition, $y_{(i,j)}$ denotes the sign of the link $e_{ij}$, where $y_{(i,j)}=1$ represents that $u_i$ and $u_j$ are friends and $y_{(i,j)}=-1$ represents that $u_i$ and $u_j$ will not build a link between each other. Then, we can use this equation to update the community assignments iteratively.

{\bf Hierarchical Priors} Different from the original MMSB in~\cite{Airoldi:2008:MMS:1390681.1442798}, we need to derive two priors: $\lambda_d$ for each network and $\mathbf{x}_i$ for each user $u_i$. To find the optimal values, we first define the union distribution over community assignments and priors as
\begin{eqnarray}
&&p(\mathbf{z},\lambda,\mathbf{x})=\nonumber\\
&&\Pi_{d,i}\frac{\Gamma(\sum_k{t(\mathbf{x}_i\lambda_{d,k}^T)})}{\Gamma(\sum_k{t(\mathbf{x}_i\lambda_{d,k})}+n_{i,\cdot}^d)}\Pi_{k}\frac{\Gamma(t(\mathbf{x}_i\lambda_{d,k}^T)+n^d_{i, k})}{\Gamma(t(\mathbf{x}_i\lambda_{d,k}^T))}\nonumber\\
&&\Pi_{d,k,k',y}\frac{\Gamma(t(\lambda_d^T\lambda_d)_{k,k'}+1+n_{d,k,k',y})}{\Gamma(t(\lambda_d\lambda_d^T)_{k,k'}+1)}\nonumber\\
&&\Pi_{d,k,t}\frac{1}{2\pi\sigma_d^2}t\Big(-\frac{\lambda^2_{d,k,t}}{2\sigma_d^2}\Big)\Pi_{i,t}\frac{1}{2\pi\sigma_u^2}t\Big(-\frac{x^2_{i,t}}{2\sigma_u^2}\Big)
\end{eqnarray}
Note that this equation is similar to Eq.(4) in~\cite{Low:2011:MDU:2020408.2020434} and Eq.(1) in~\cite{DBLP:conf/uai/MimnoM08}, but with very different terms, since we apply $\lambda_d$ as a prior to the compact matrix $\mathbf{B}_d$. Let $dt(x)=\partial_xt(x)$ be the derivative of the transform function. The derivative of the log of likelihood with respect to $\lambda_{d,k,t}$ for a given community $k$ and the feature $t$ is
\begin{eqnarray}
\label{eq:lbfgs}
&&\partial_{\lambda_{d,k,t}}\log{p(\mathbf{z},\lambda,\mathbf{x})}=-\frac{\lambda_{d,k,t}}{\sigma^2_d}+\sum_{d,i}x_{i,t}dt(\mathbf{x}_i\lambda_{d,k}^T)\nonumber\\
&&\Big[\Psi\Big(\sum_{k}dt(\mathbf{x}_i\lambda_{d,k}^T)\Big)-\Psi\Big(\sum_{k}dt(\mathbf{x}_i\lambda_{d,k}^T)+n^d_{i,\cdot}\Big)\nonumber\\
&&+\Psi\big(dt(\mathbf{x}_i\lambda_{d,k}^T)+n^d_{i,t}\big)-\Psi\big(dt(\mathbf{x}_i\lambda_{d,k}^T)\big)\Big]\nonumber\\
&&+\sum_{d,k,k',y}\lambda_{d,k',t}dt(\lambda_{d,k}\lambda_{d,k'}^T)\nonumber\\
&&\Big[\Psi\big(dt(\lambda_{d,k}\lambda_{d,k'}^T)+n^d_{k,k',y}\big)-\Psi\big(dt(\lambda_{d,k}\lambda_{d,k'}^T)\big)\Big]
\end{eqnarray}
We exploit a standard L-BFGS optimizer~\cite{Liu:1989:LMB:81100.83726} using the above equation to update each $\lambda_d$. Similarly, the derivative of the log of likelihood with respect to the parameter $x_{i,t}$ is similar:
\begin{eqnarray}
&&\partial_{x_{i,t}}\log{p(\mathbf{z},\lambda,\mathbf{x})}=-\frac{x_{i,t}}{\sigma^2_u}+\sum_{d,i}\sum_{k}\lambda_{d,k,t}dt(\mathbf{x}_i\lambda_{d,k}^T)\nonumber\\
&&\Big[\Psi\Big(\sum_{k}dt(\mathbf{x}_i\lambda_{d,k}^T)\Big)-\Psi\Big(\sum_{k}dt(\mathbf{x}_i\lambda_{d,k}^T)+n^d_{i,\cdot}\Big)\nonumber\\
&&+\Psi\big(dt(\mathbf{x}_i\lambda_{d,k}^T)+n^d_{i,t}\big)-\Psi\big(dt(\mathbf{x}_i\lambda_{d,k}^T)\big)\Big]
\end{eqnarray}
We adopt a Metropolis-Hastings sampler~\cite{Robert:2005:MCS:1051451} to search an optimal $\mathbf{x}_i$ for each user $u_i$. Suppose the current value is $x_{i,t}$, we propose the new value:
\begin{eqnarray}
\bar{x}_{i,t}=x_{i,t}+\frac{\sigma^2}{2}\frac{\partial\log{p(\mathbf{z},\lambda,\mathbf{x})}}{\partial{x_{i,t}}}{\mid}_{x_{i,t}}
\end{eqnarray}
Then, we compute the acceptance ratio as
\begin{eqnarray}
\label{eq:ratio}
r=\frac{\exp\Big\{-\parallel{x_{i,t}-\bar{x}_{i,t}-\frac{\sigma^2}{2}\frac{\partial\log{p(\mathbf{z},\lambda,\mathbf{x})}}{\partial{x_{i,t}}}{\mid}_{\bar{x}_{i,t}}}\parallel/2\sigma^2\Big\}}{\exp\Big\{-\parallel{\bar{x}_{i,t}-x_{i,t}-\frac{\sigma^2}{2}\frac{\partial\log{p(\mathbf{z},\lambda,\mathbf{x})}}{\partial{x_{i,t}}}{\mid}_{x_{i,t}}}\parallel/2\sigma^2\Big\}}
\end{eqnarray}
where $\sigma$ is related to the learning rate. Finally, we update $x_{i,t}$ to the new value $\bar{x}_{i,t}$ with probability $\min(r,1)$.

\paragraph{Inference Framework} The complete inference process is described in Algorithm~\ref{algo:MMSB}. Generally, we alternatively optimize the user and network priors, and latent variables in MMSB as well as the community assignments. Specifically, we randomly initialize the priors according to the assigned distributions. Then we use these priors to sample the latent variables of MMSB, including $\pi$ and $\mathbf{B}$. Consequently, we alternatively update the community assignments and latent variables in each MMSB model, and the network-specific and user-dependent priors. This process is repeated until convergence. To avoid overfitting and reduce the computational cost, we update both $\lambda$ and $\mathbf{x}$ every 10 iterations and set the number of iterations for L-BFGS as 10.

At each iteration, Gibbs sampling needs to look up all $m$ links in nested networks and then update $\pi$ with time $O(NKn)$ where $N$ is the number of networks and $n$ is the number of users. In addition, L-BFGS takes $O(NKT)$ to update $\lambda$ and Metropolis-Hastings sampler spends $O(Tn)$ on updating $\mathbf{x}$. Typically, $NKT$ is much smaller than $NKn$. Thus, with $I$ iterations, the time complexity is $O(I(m+Tn+NKn))$ which linearly increases with the number of links $m$ and and number of users $n$.

\begin{algorithm}[t]
{ \caption{\small Inference for ComFP} \label{algo:MMSB}
\begin{algorithmic}[1]
\begin{small}
    \STATE {\bf Input}: Nested social networks: $\mathbb{G}$, Number of iterations $I$, Number of user features $T$ and Number of communities $K$
    \STATE {\bf Output}: Compact matrix $\mathbf{B}_d$ and membership vector $\pi_{id}$
    \STATE Generate $\lambda_d$ and $\mathbf{x}_i$ randomly
    \STATE Initialize the compact matrices and membership vectors
    \FOR {$k=1$ to $I$}
        \STATE Perform Gibbs sampling using Eq.(\ref{eq:gibbs})
        \STATE Update membership vector $\pi_{id}$ and compact matrix $\mathbf{B}_d$
        \STATE {\bf IF} {k\%10==0}
        \STATE \quad\quad Update $\lambda_d$ using L-BFGS using Eq.(\ref{eq:lbfgs})
        \STATE \quad\quad Update $\mathbf{x}_i$ using Metropolis-Hastings in Eq.(\ref{eq:ratio})
        \STATE {\bf IF} {the algorithm is convergent} {\bf Break}
    \ENDFOR
    \STATE {\bf Return} $\mathbf{B}_d$ and $\pi_{id}$
\end{small}
\end{algorithmic}}
\end{algorithm}

\begin{table*}[t]
\caption{\small Summary of Data Characteristics}
\begin{center}
\begin{scriptsize}
\begin{tabular}{|c||c|c|c|c|l|}
  \hline
  {Collections} & \#User & \#Relations/Interactions & Types of Relations/Interactions \\
  \hline
  \hline
  Tencent  & $\sim{1M}$    & $\sim{110M}$ & Instant Messaging (QQ), Microblog Following (MB) \\
  \hline
  Douban   & $\sim{0.05M}$   & $\sim{1M}$   & Online, Offline \\
  \hline
  Epinion     & $\sim{0.1M}$ & $\sim{0.8M}$ & Trust, Distrust \\
  \hline\hline
  Facebook & $\sim{0.06M}$ & $\sim{1.8M}$ & Link, Wall Posting \\
  \hline
  Renren   & $\sim{0.5M}$  & $\sim{32M}$  & Footprint, Visiting, Talking, Buddy Application \\
  \hline
  Twitter  & $\sim{0.3M}$  & $\sim{0.9M}$ & Forwarding, Mention \\
  \hline
  Sina Weibo  & $\sim{6M}$ & $\sim{320M}$ & Forwarding, Mention \\
  \hline
  Github      & $\sim{0.05M}$ & $\sim{1M}$ & Following, Collaborating \\
  \hline
  StackOverflow (SO)  & $\sim{0.8M}$ & $\sim{33M}$ & Answering, Commenting, Voting \\
  \hline
\end{tabular}
\label{tab:network}
\end{scriptsize}
\end{center}
\vskip -0.2in
\end{table*}

\begin{figure*}[t]
\begin{small}
\centering \mbox{
\subfigure[Tencent]{\scalebox{0.5}{\includegraphics[width=\columnwidth]{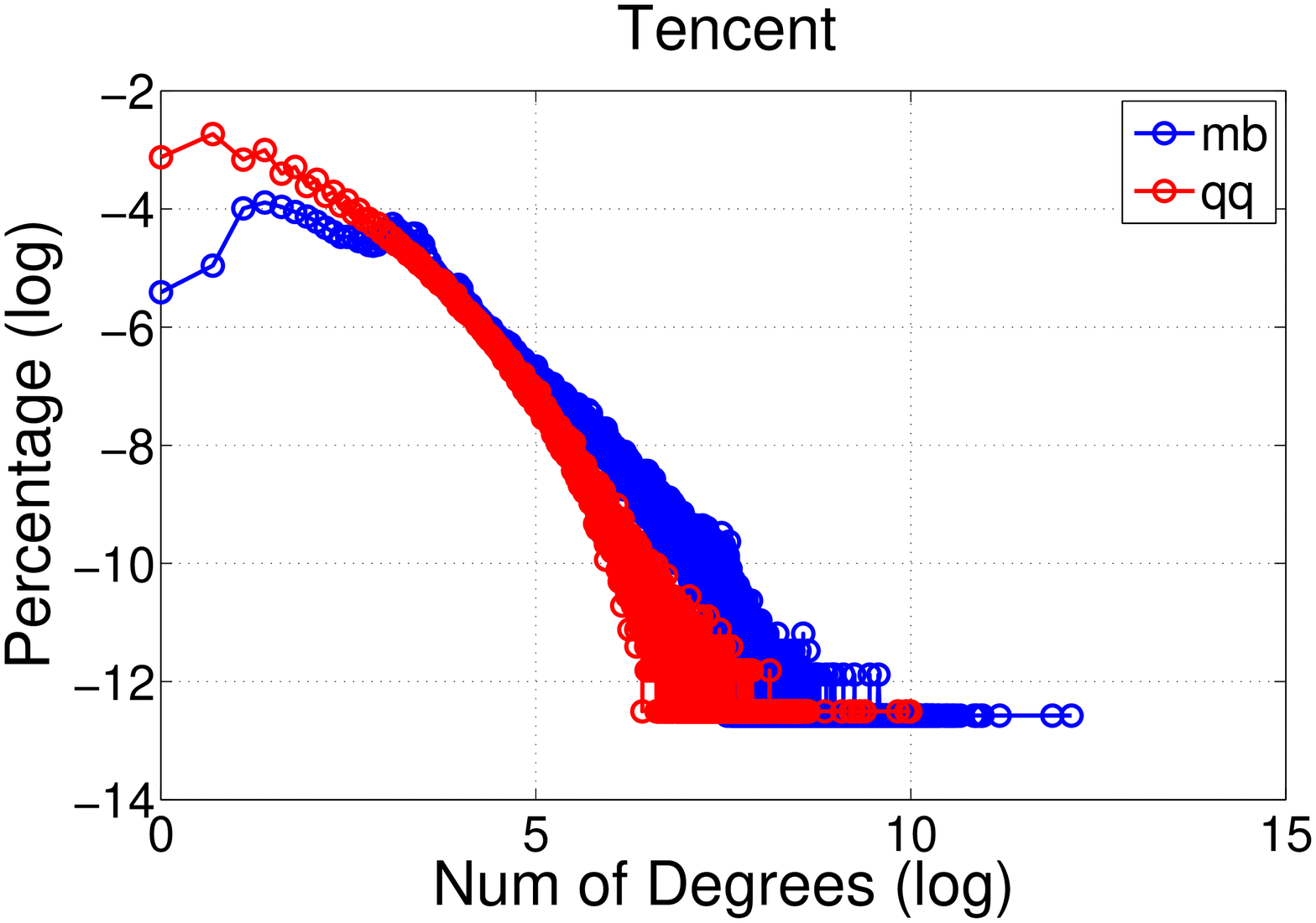}}}
\subfigure[Douban]{\scalebox{0.5}{\includegraphics[width=\columnwidth]{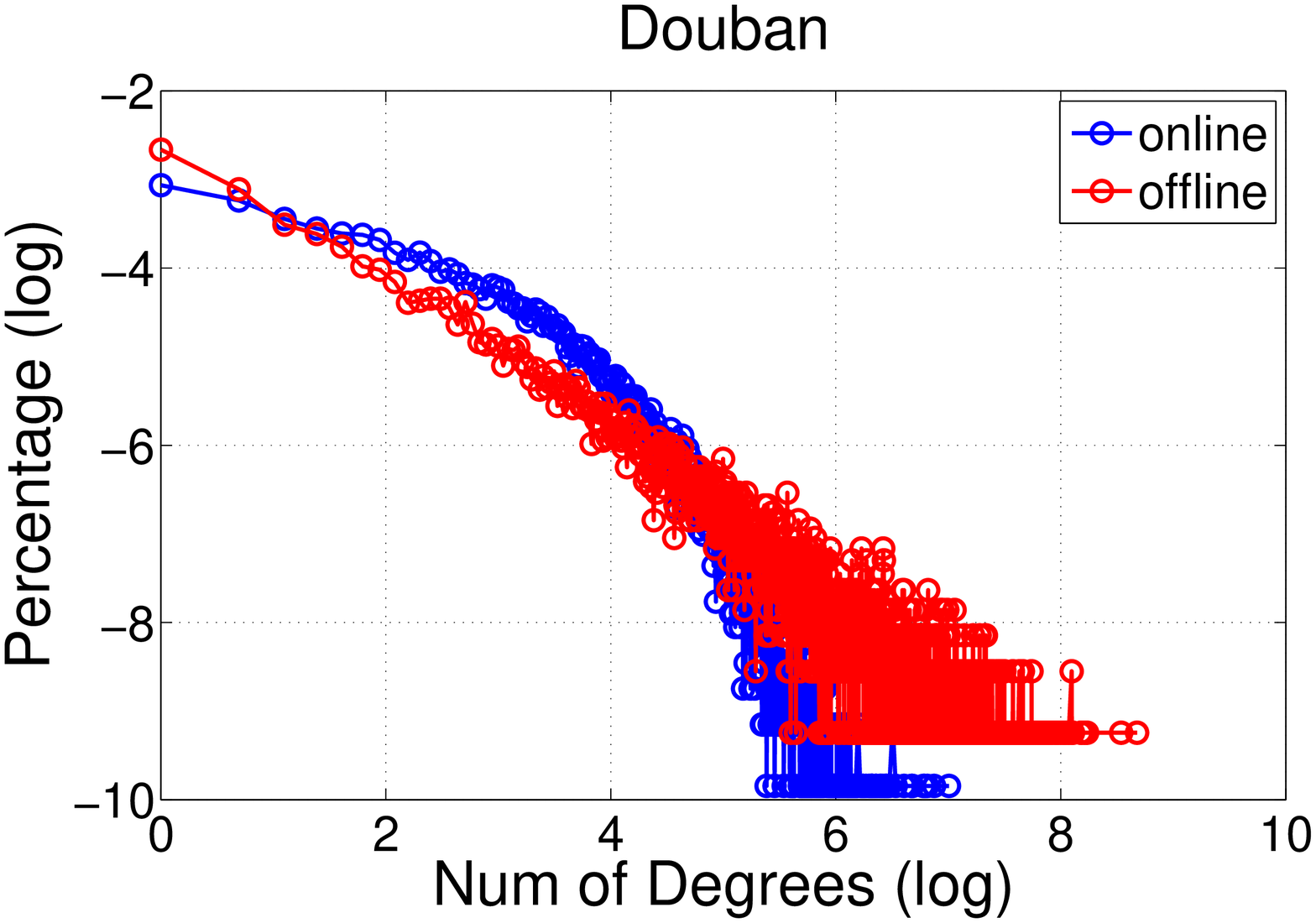}}}
\subfigure[Epinion]{\scalebox{0.5}{\includegraphics[width=\columnwidth]{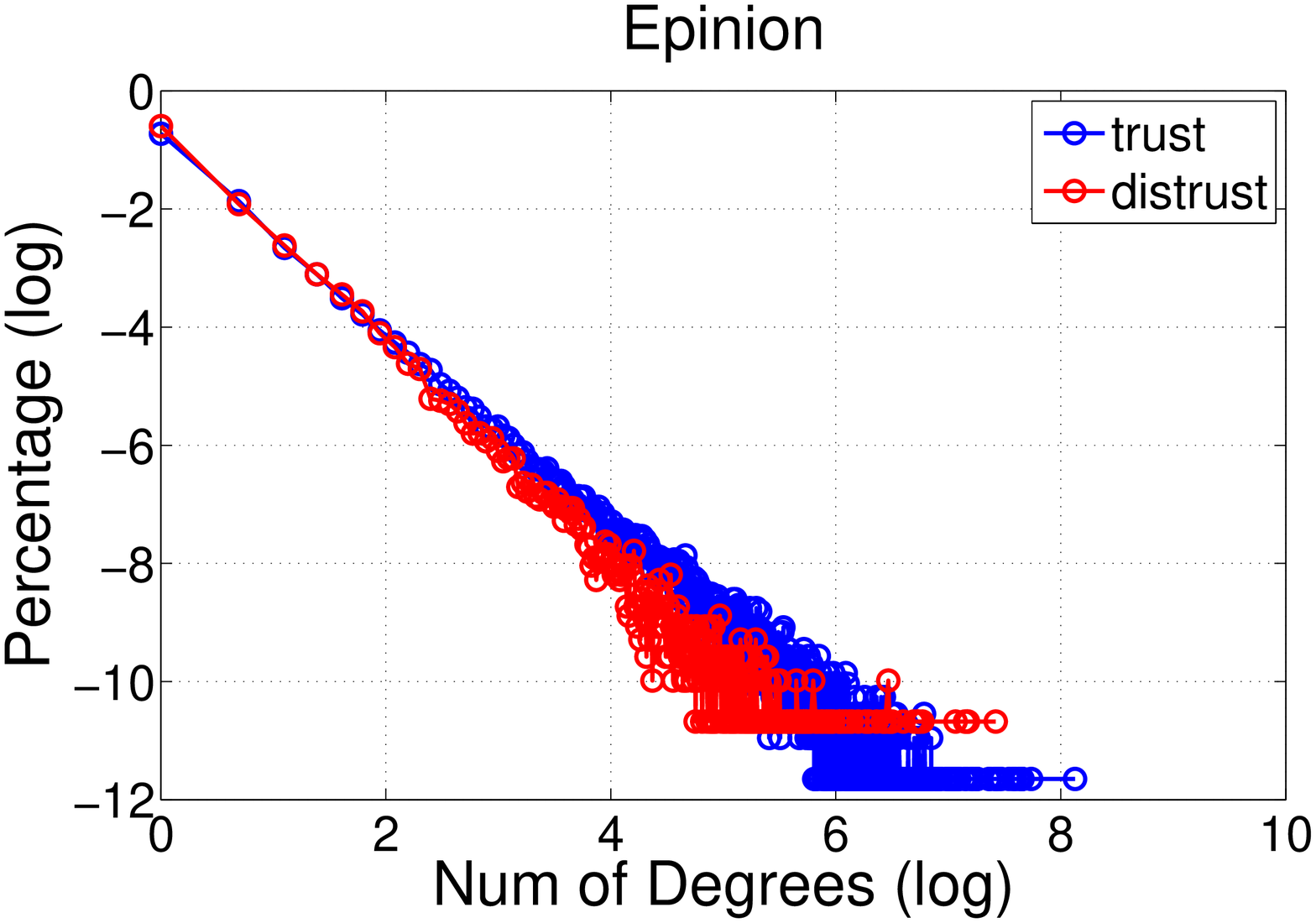}}}
}
\mbox{
\subfigure[Facebook]{\scalebox{0.5}{\includegraphics[width=\columnwidth]{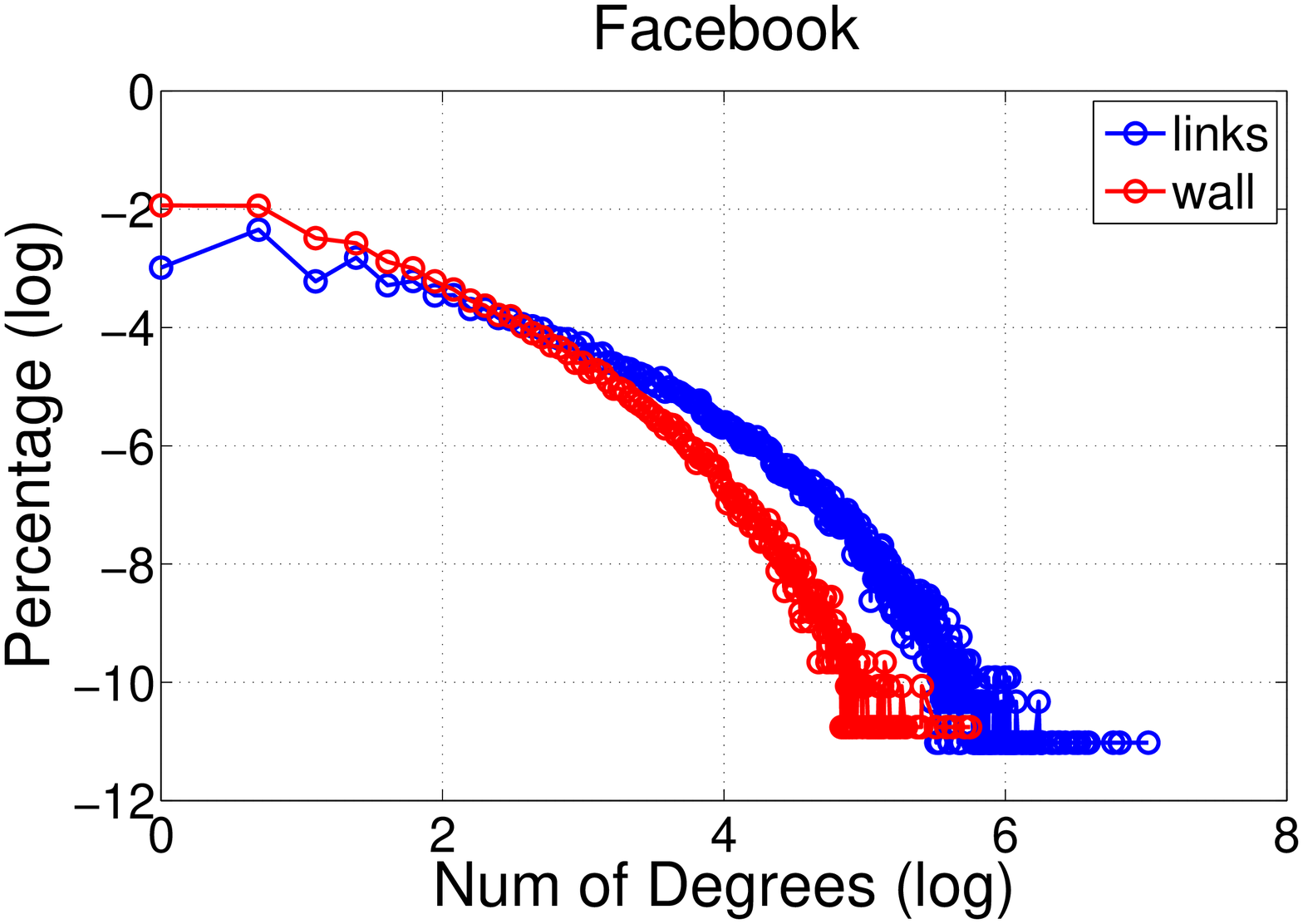}}}
\subfigure[Renren]{\scalebox{0.5}{\includegraphics[width=\columnwidth]{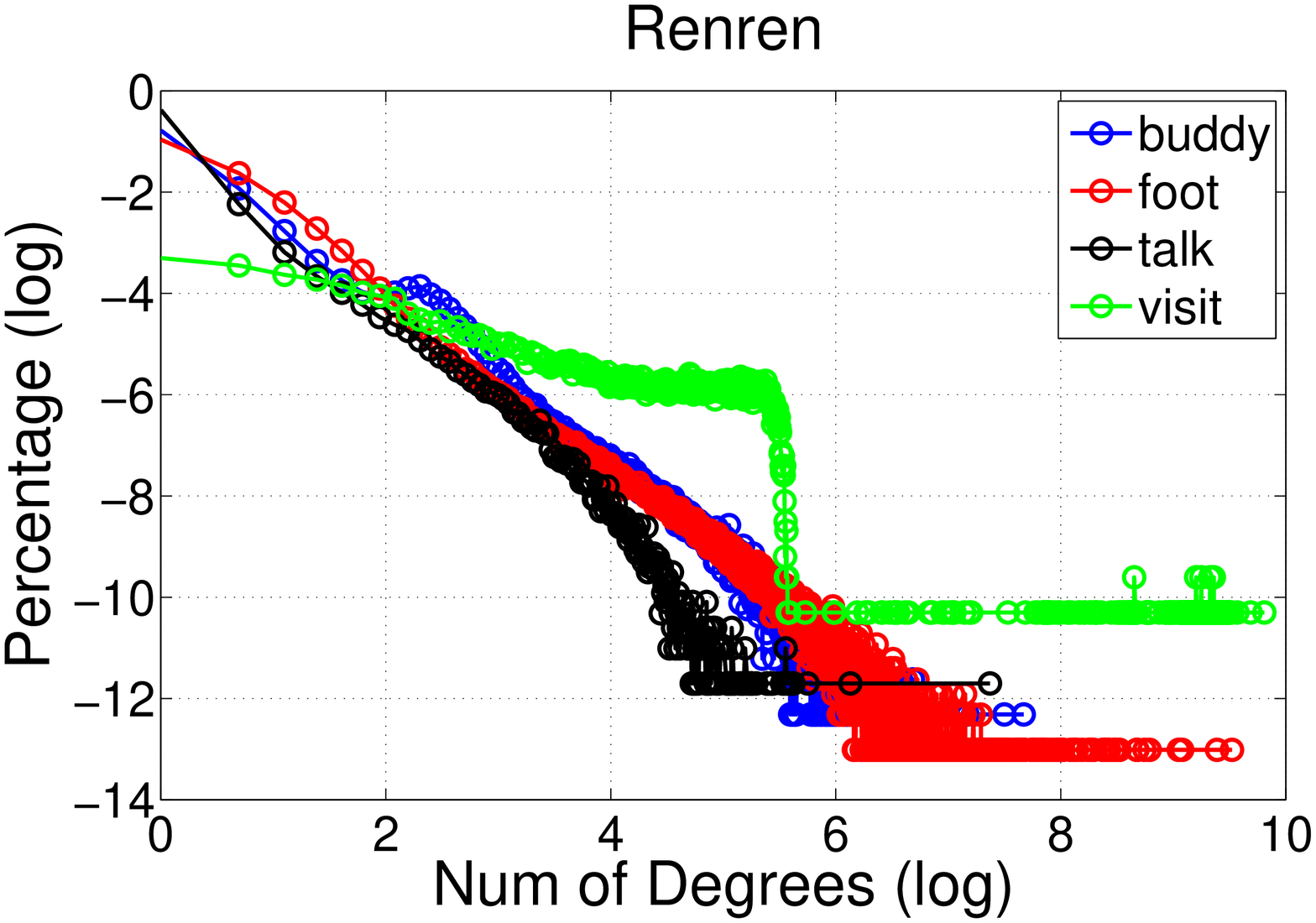}}}
\subfigure[Twitter]{\scalebox{0.5}{\includegraphics[width=\columnwidth]{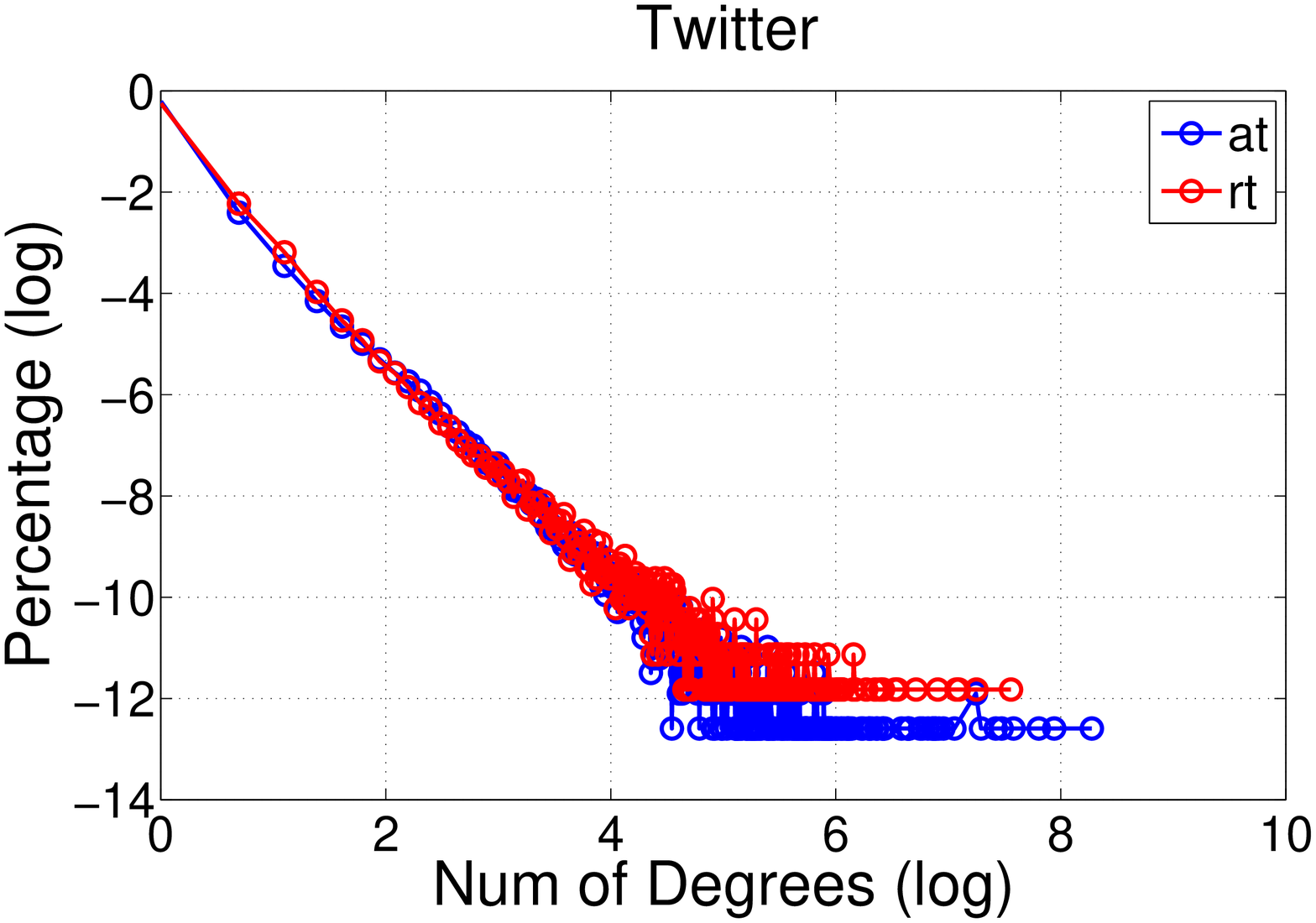}}}
}
\mbox{
\subfigure[Weibo]{\scalebox{0.5}{\includegraphics[width=\columnwidth]{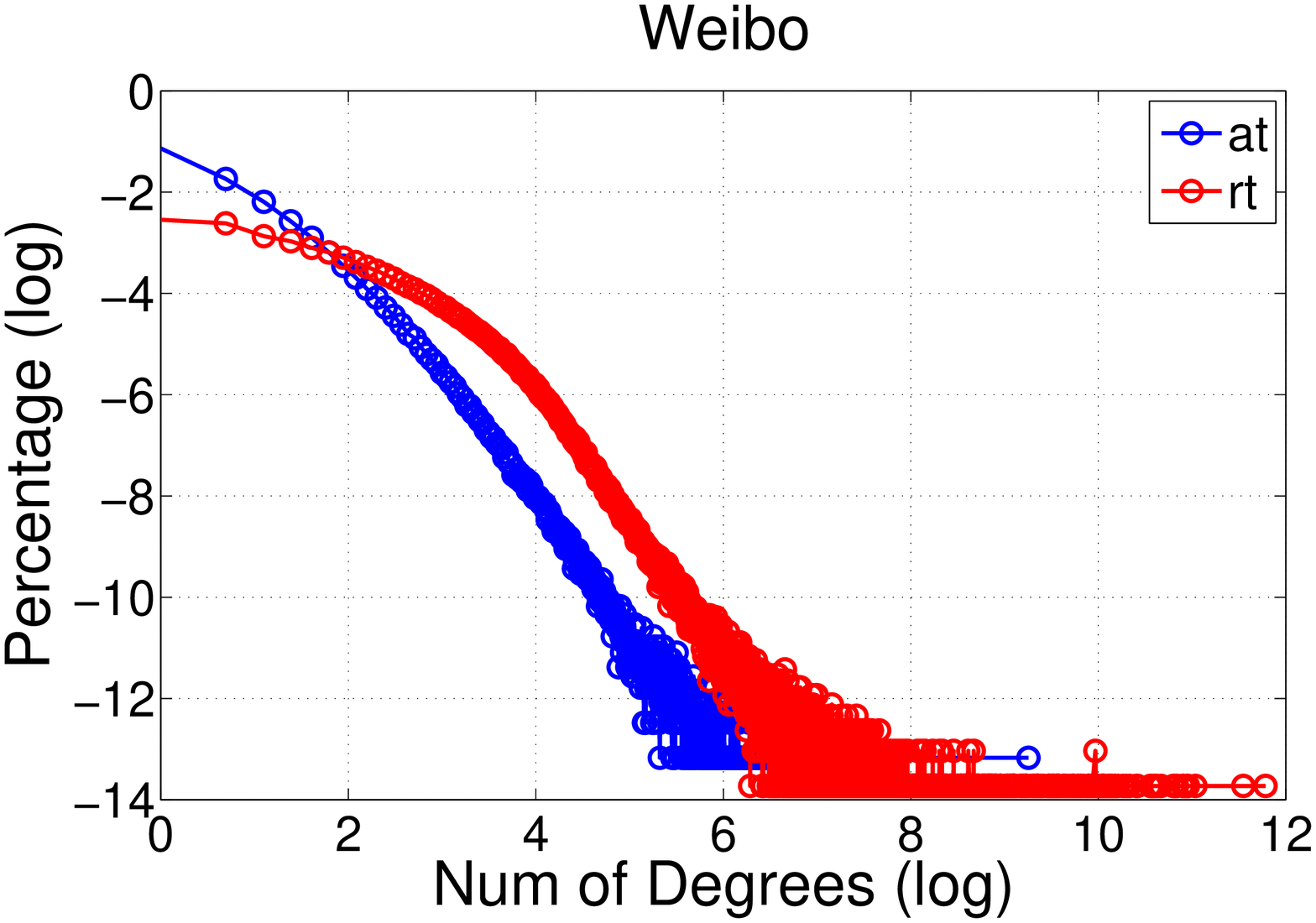}}}
\subfigure[Github]{\scalebox{0.5}{\includegraphics[width=\columnwidth]{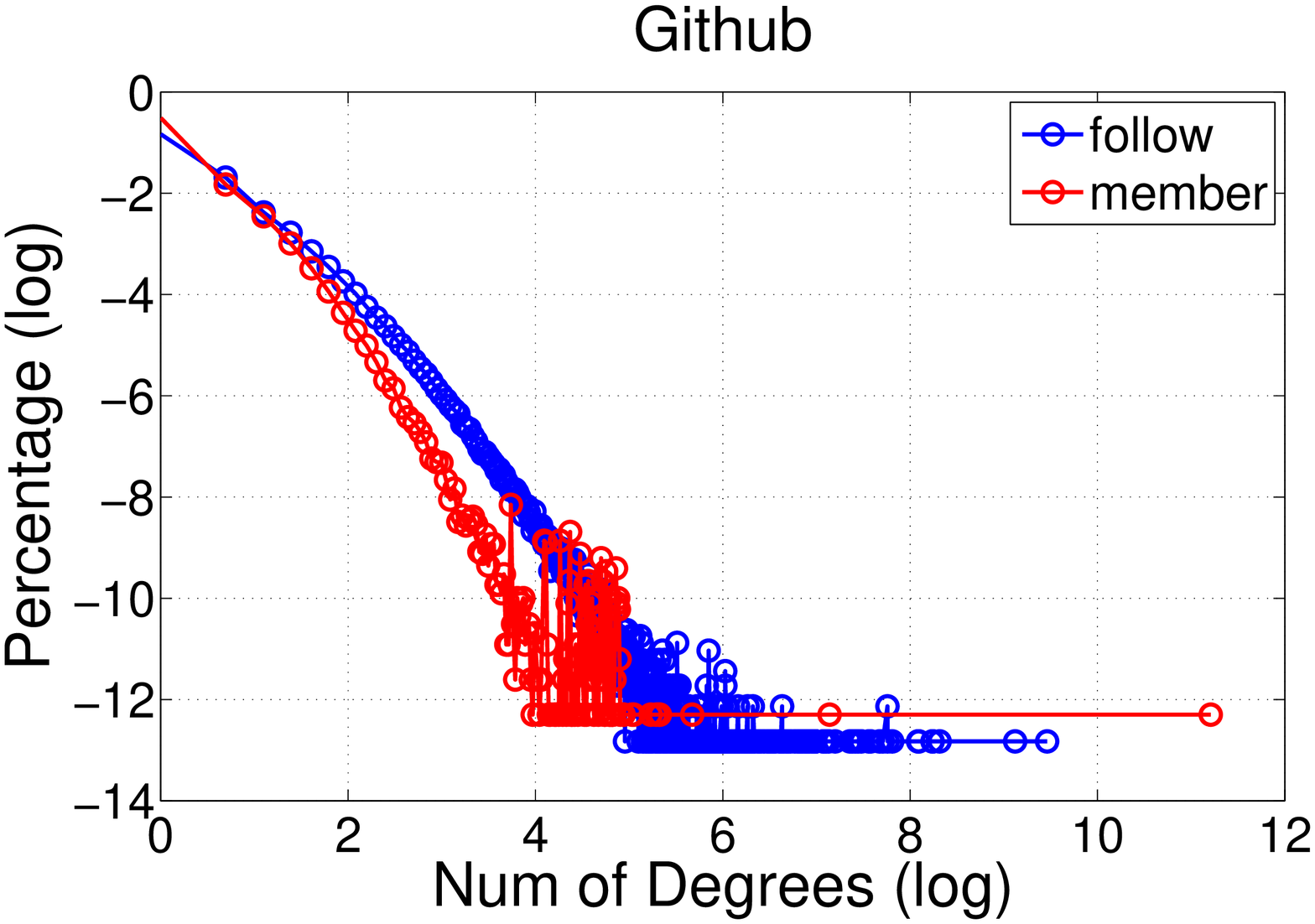}}}
\subfigure[StackOverflow]{\scalebox{0.5}{\includegraphics[width=\columnwidth]{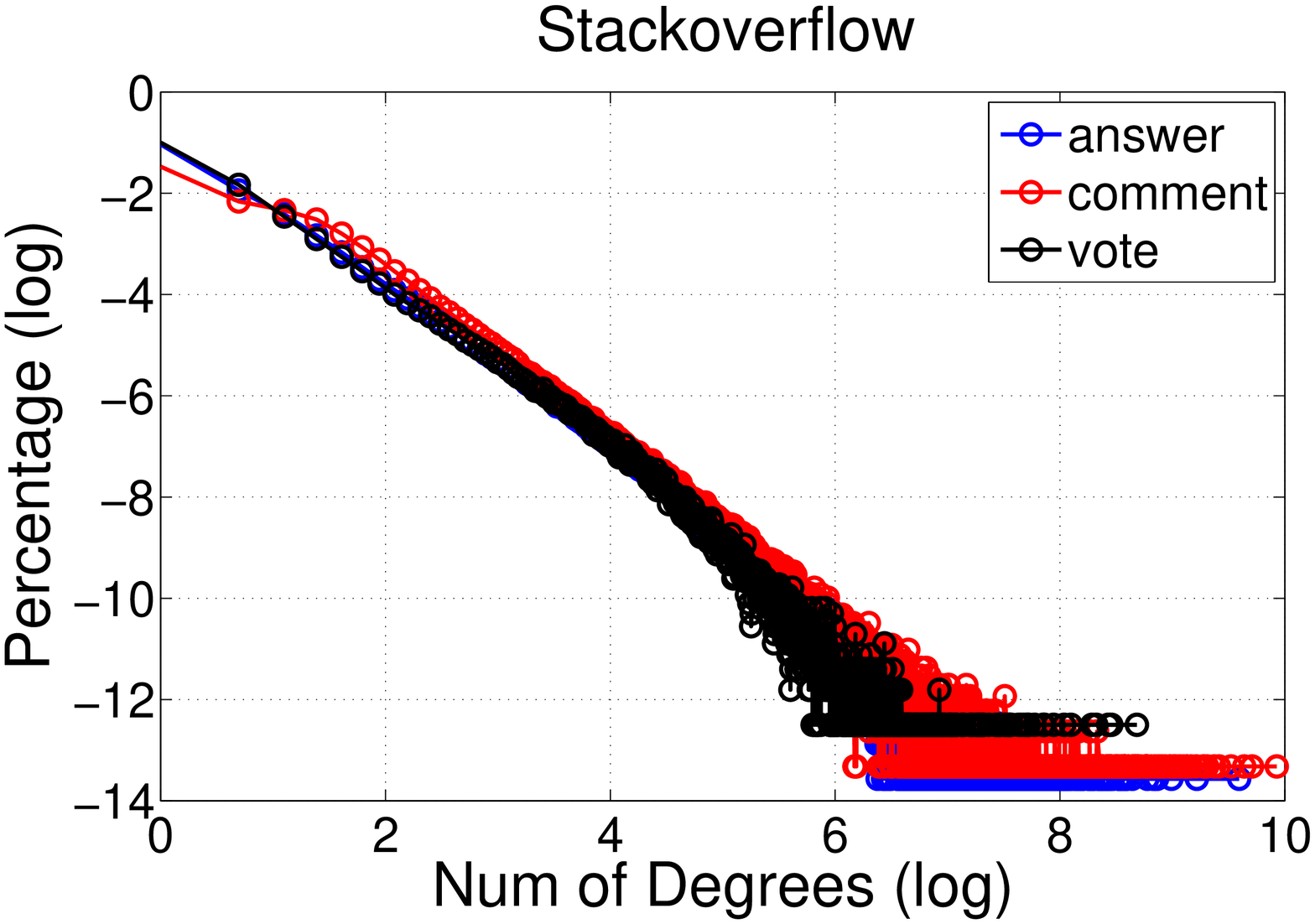}}}
}\vskip -0.05in
\end{small}
\caption{Degree Distributions} \label{fig:deg}
\vskip -0.1in
\end{figure*}

\section{Data Description}
We evaluate the effectiveness of the proposed method on nine real-world datasets from Tencent\footnote{\url{http://www.tencent.com/en-us/index.shtml}}, Douban\footnote{\url{http://www.douban.com}}, Sina Weibo\footnote{\url{http://www.wise2012.cs.ucy.ac.cy/challenge.html}}, Renren\footnote{\url{http://www.renren.com}}, Facebook\footnote{\url{http://socialnetworks.mpi-sws.org}}, Twitter~\cite{icwsm12kooti}, Github\footnote{\url{http://www.githubarchive.org/}}, Stackoverflow\footnote{\url{http://meta.stackoverflow.com/}} and Epinion\footnote{\url{http://konect.uni-koblenz.de/networks/epinions}}. According to different link types, these datasets can be classified as relational networks, e.g., Tencent, Douban and Epinion, and interaction networks (the remaining six datasets). User pairs in relational networks are distinct but users can interact with each other multiple times in interaction networks. Specifically, Tencent collection contains an instant messaging network (QQ) and a Microblog network (MB) and Douban collection contains users' online and offline relationships. In Epinion, people can build trust or distrust relations with others. Facebook collection captures users' friendships and wall posting actions. In Renren collection, users can leave footprint, visit friends' homepages, talk with friends, and send buddy applications. For Twitter and Sina Weibo, people can forward (RT) and mention (@) others' tweets. In Github, users follow each other as well as being collaborators in different projects. In Stackoverflow, people can answer others' questions, comment on others' answers and vote others' posts.

Data statistics can be found in Table~\ref{tab:network}. Their degree distributions are plotted in Figure~\ref{fig:deg}. Although the degree distributions all follow the power-low distribution, different individual networks in each collection have similar but different properties. It also implies that simply merging networks does not work, where networks' specific patterns can be hidden. In summary, these datasets come from different applications with different scales. The correlations between individual networks are very different. For example, forwarding and mention may have highly positive correlation while trust and distrust relations are strongly negative correlated. This variety property makes the experiments convincing. To crawl data, we employ random walk based sampling method to select sub networks in Tencent, Douban and Renren networks and extract relational knowledge from whole public data dumps of other datasets. In each dataset, users in different individual networks can be identified by unified user identity, such as the QQ number in the Tencent collection. In addition, some recent techniques can also solve this user alignment problem~\cite{Liu:2013:WNU:2433396.2433457,Yuan:2013:MSRA,Liu:2013:CM}.

%
\begin{table*}[t]
\caption{Performance Comparisons}
\begin{center}
\begin{small}
\begin{tabular}{|c|c||c|c|c|c|c|}
  \hline
  \multicolumn{2}{|c||}{Networks} & MMSB & MMSB-C & TF & MRLP & ComFP   \\
  \hline
  \multirow{2}{*}{Tencent} & \mbox{QQ} & 0.4885 & 0.4821 & 0.5011 & 0.5034 & 0.5473\\
  & \mbox{MB} & 0.2053 & 0.2042 & 0.2351 & 0.2387 & 0.2521 \\
  \hline
  \multirow{2}{*}{Douban} & \mbox{Offline} & 0.7213 & 0.7188 & 0.7312 & 0.7343 & 0.7521\\
  & \mbox{Online} & 0.6346 & 0.6510 & 0.6541 & 0.6567 & 0.6793 \\
  \hline
  \multirow{2}{*}{Epinion} & \mbox{Trust} & 0.6831 & 0.6527 & 0.6923 & 0.6933 & 0.7312\\
  & \mbox{Distrust} & 0.6629 & 0.6314 & 0.6679 & 0.6687 & 0.7089\\
  \hline\hline
  \multirow{2}{*}{Facebook} & \mbox{Link} & 0.7078 & 0.7107 & 0.7143 & 0.7166 & 0.7365\\
  & \mbox{Wall} & 0.6681 & 0.6567 & 0.6761 & 0.6781 & 0.6939\\
  \hline
  \multirow{4}{*}{Renren} & \mbox{Footprint} & 0.4912 & 0.5088 & 0.4981 & 0.5088 & 0.5435\\
  & \mbox{Visiting} & 0.1932 & 0.1965 & 0.2045 & 0.2145 & 0.2431\\
  & \mbox{Talk} & 0.2621 & 0.2459 & 0.2777 & 0.2875 & 0.3189\\
  & \mbox{Buddy} & 0.3388 & 0.3992 & 0.4055 & 0.4027 & 0.4481\\
  \hline
  \multirow{2}{*}{Twitter} & \mbox{Forwarding} & 0.6613 & 0.6722 & 0.6741 & 0.6734 & 0.6923\\
  & \mbox{Mention} & 0.7432 & 0.7312 & 0.7686 & 0.7742 & 0.7921\\
  \hline
  \multirow{2}{*}{Weibo} & \mbox{Forwarding} & 0.5357 & 0.5351 & 0.5713 & 0.5817 & 0.6101\\
  & \mbox{Mention} & 0.6489 & 0.6551 & 0.6715 & 0.6887 & 0.7011\\
  \hline
  \multirow{2}{*}{Github} & \mbox{Following} & 0.7236 & 0.7343 & 0.7414 & 0.7501 & 0.7901\\
  & \mbox{Collaborating} & 0.7911 & 0.7921 & 0.8065 & 0.8078 & 0.8265\\
  \hline
  \multirow{3}{*}{SO} & \mbox{Answering} & 0.8101 & 0.8132 & 0.8288 & 0.8298 & 0.8532 \\
  & \mbox{Commenting} & 0.7576 & 0.7554 & 0.8076 & 0.8012 & 0.8189 \\
  & \mbox{Voting} & 0.7732 & 0.7701 & 0.7965 & 0.8054 & 0.8187 \\
  \hline
\end{tabular}
\label{tab:performance}
\end{small}
\end{center}
\vskip -0.2in
\end{table*}

\section{Experiments}
We show that the proposed algorithm is better than merging all networks simply, and better than only considering single networks. Thus, we introduce two baselines based on MMSB. One is to learn an MMSB on each individual network and the other is C-MMSB which performs MMSB on a combined network, of which the edge set is the union of all edge sets from all individual networks. In addition, we introduce two other baselines that are applied on multi-relational networks: TF~\cite{DBLP:conf/cidm/GaoDG11} and MRLP~\cite{Davis:2011:MLP:2055438.2055676}, where TF formulates multiple relationships as a tensor and then performs factorization and MRLP performs link prediction based on triangle patterns. The evaluation task is to predict which users will build links between each other. To test the performance, we select 10\% of the links in the whole dataset according to the temporal information as the hold-out set $\mathcal{T}$. In addition, to test whether ComFP can solve the data sparsity issue, we remove those popular users whose numbers of degrees are higher than the averaged degree plus one standard deviation. The numbers of user latent features and the communities in single networks are set as $25$ and we will test their effectiveness. The results are evaluated by mean average precision (MAP).
\begin{eqnarray}
MAP = \frac{1}{|\mathbf{U}|}\sum_{u\in\mathbf{U}}\frac{\sum_{(u,v)\in{\mathcal{T}_u}}\frac{1}{T}\sum_{r\in[1,T]}pre_{r}}{|\mathcal{T}_u|}
\end{eqnarray}
where $pre_r$ is the precision on top $r$ predictions. Generally, MAP measures how well the algorithms rank the links in the hold-out set against the non-existing links. As there are only positive links in the dataset, we sample non-existing links in the same magnitude randomly in the training process to construct negative examples.

\begin{figure}[t]
\centering \mbox{
\subfigure[Adaptation]{\scalebox{0.6}{\includegraphics[width=\columnwidth]{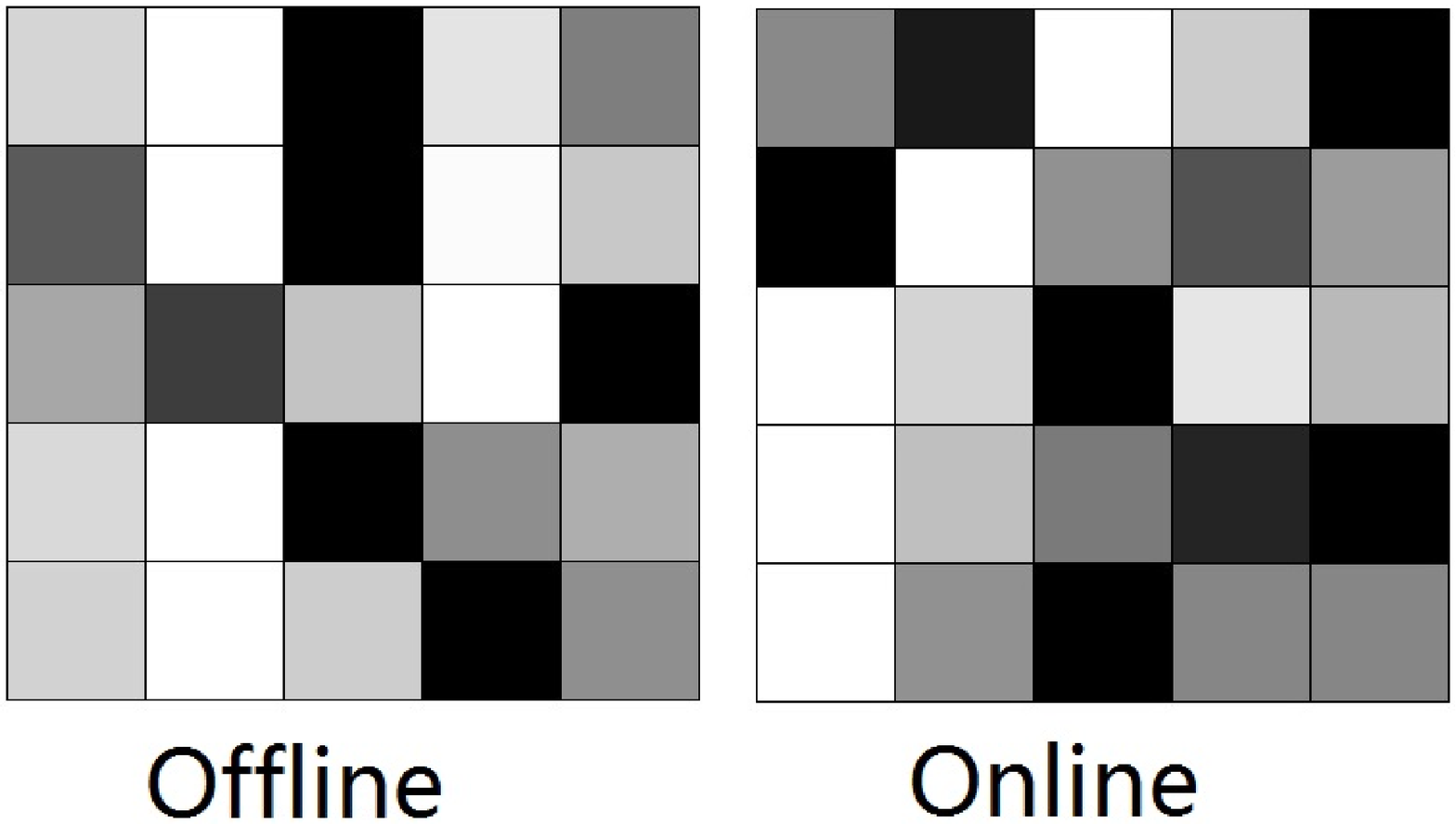}}}
}
\mbox{
\subfigure[T]{\scalebox{0.5}{\includegraphics[width=\columnwidth]{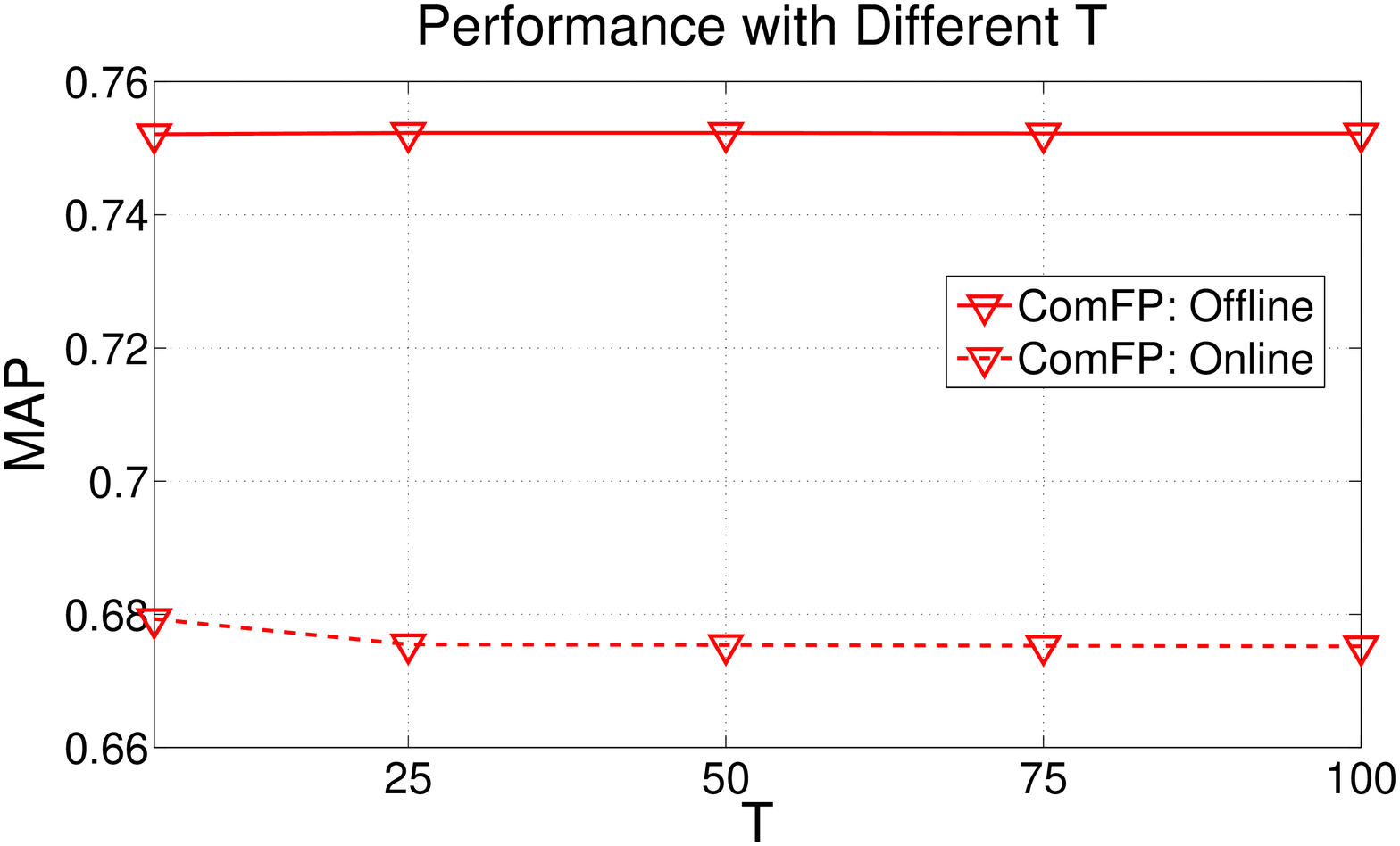}}}
\subfigure[K]{\scalebox{0.5}{\includegraphics[width=\columnwidth]{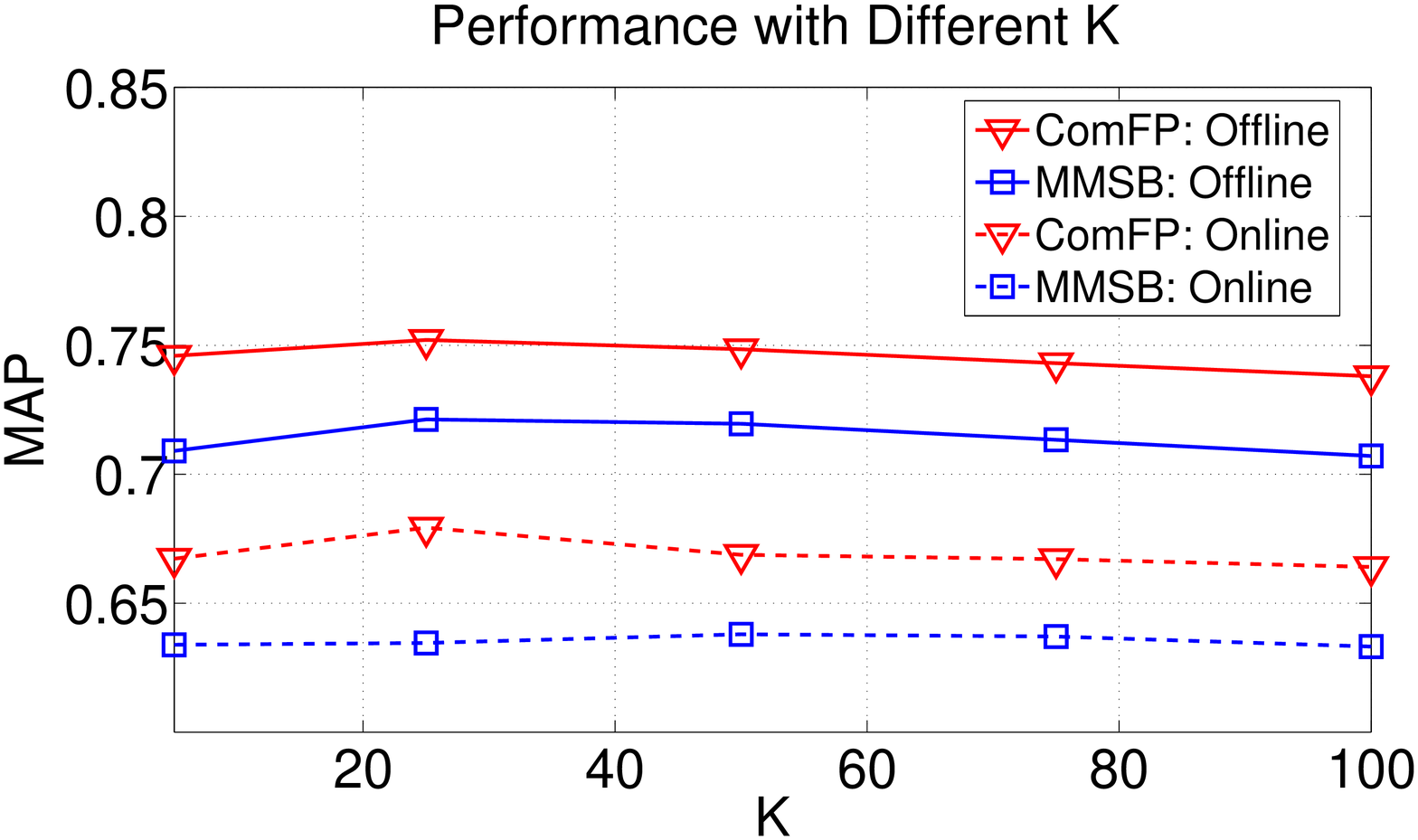}}}
}
\mbox{
\subfigure[Ratio]{\scalebox{0.5}{\includegraphics[width=\columnwidth]{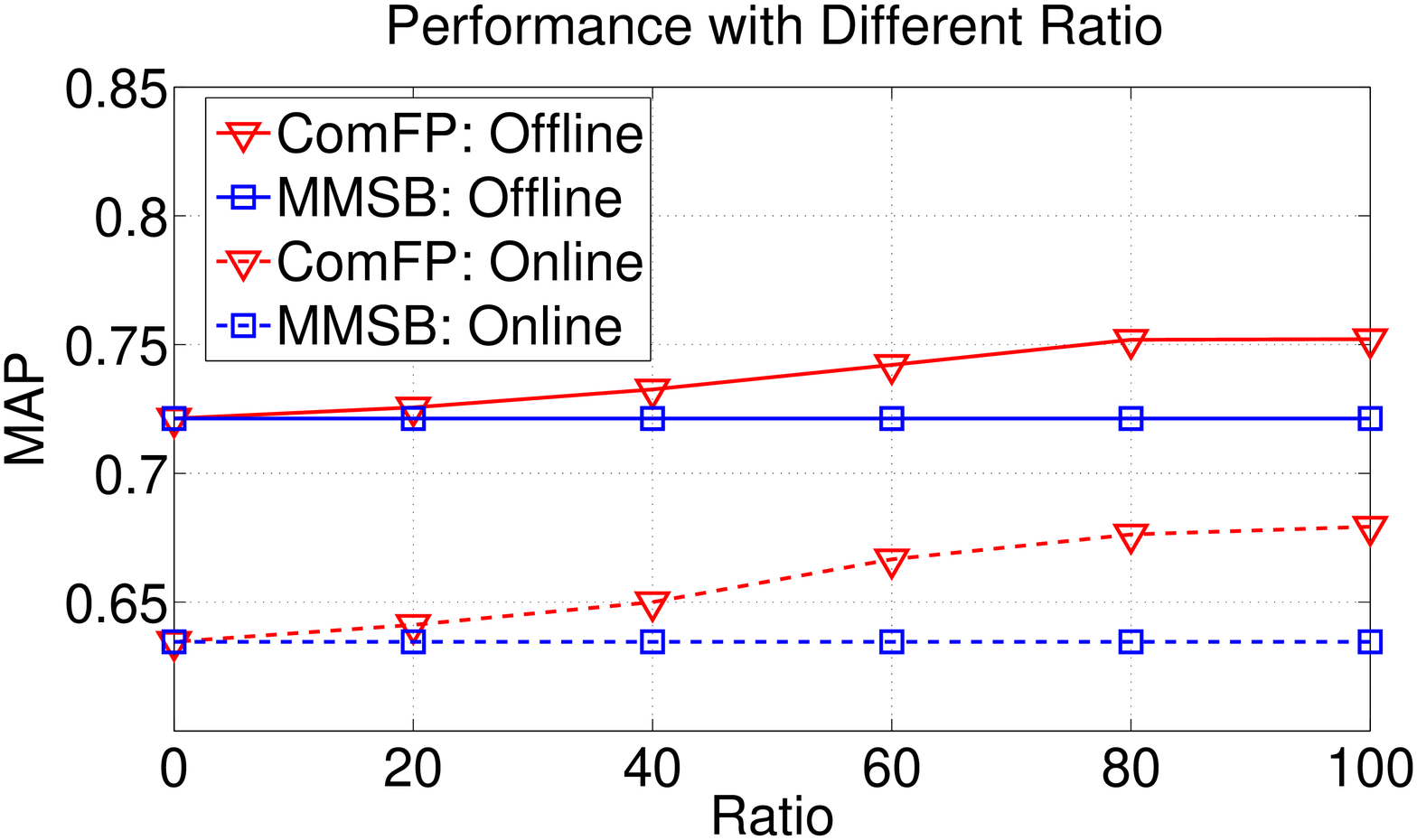}}}
\subfigure[Sparsity]{\scalebox{0.5}{\includegraphics[width=\columnwidth]{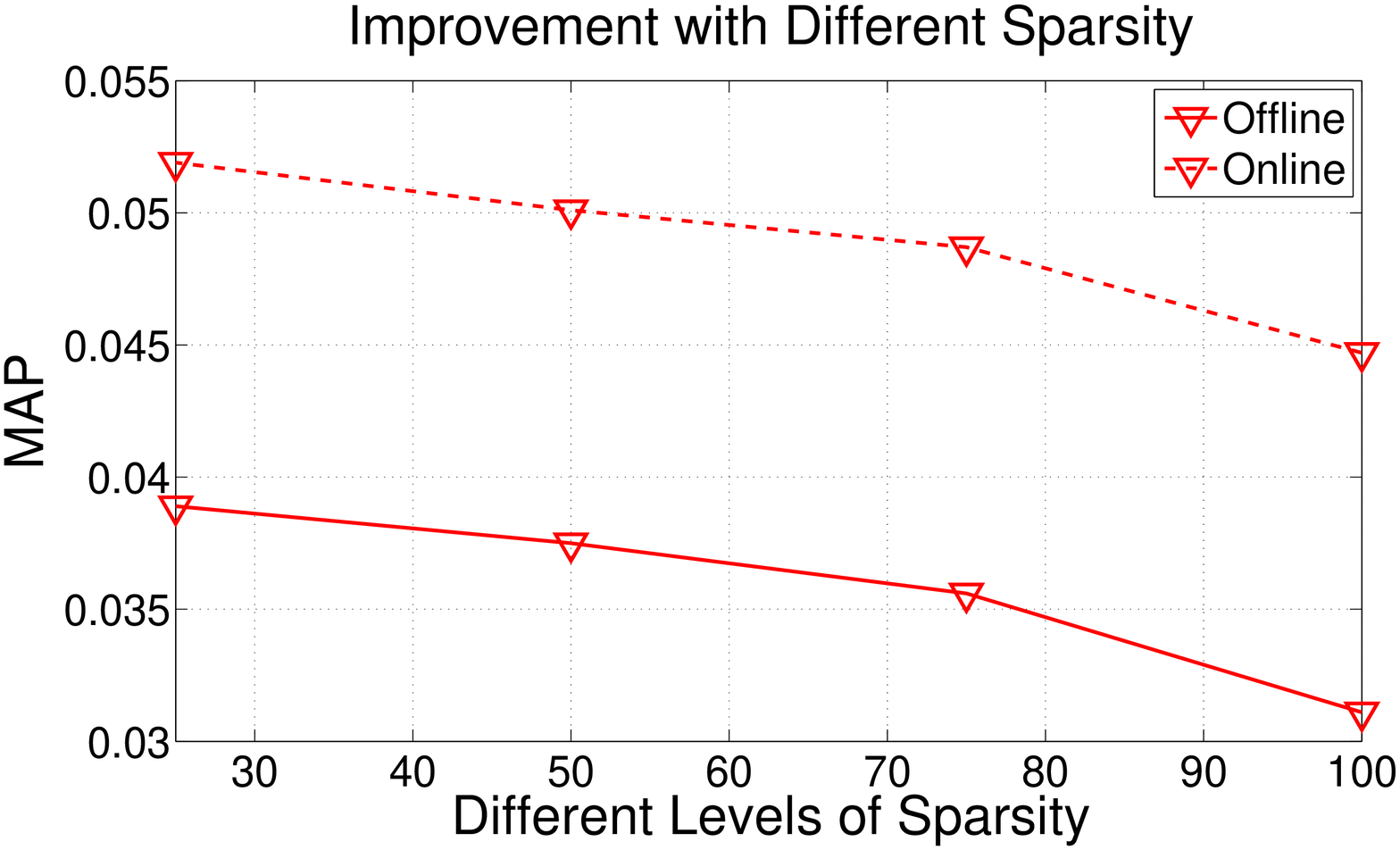}}}
}
\vskip -0.1in
\caption{Parameter Analysis on Douban collection} \label{fig:parameter}
\vskip -0.2in
\end{figure}

\begin{figure*}[t]
\begin{small}
\centering \mbox{
\subfigure[Tencent]{\scalebox{0.5}{\includegraphics[width=\columnwidth]{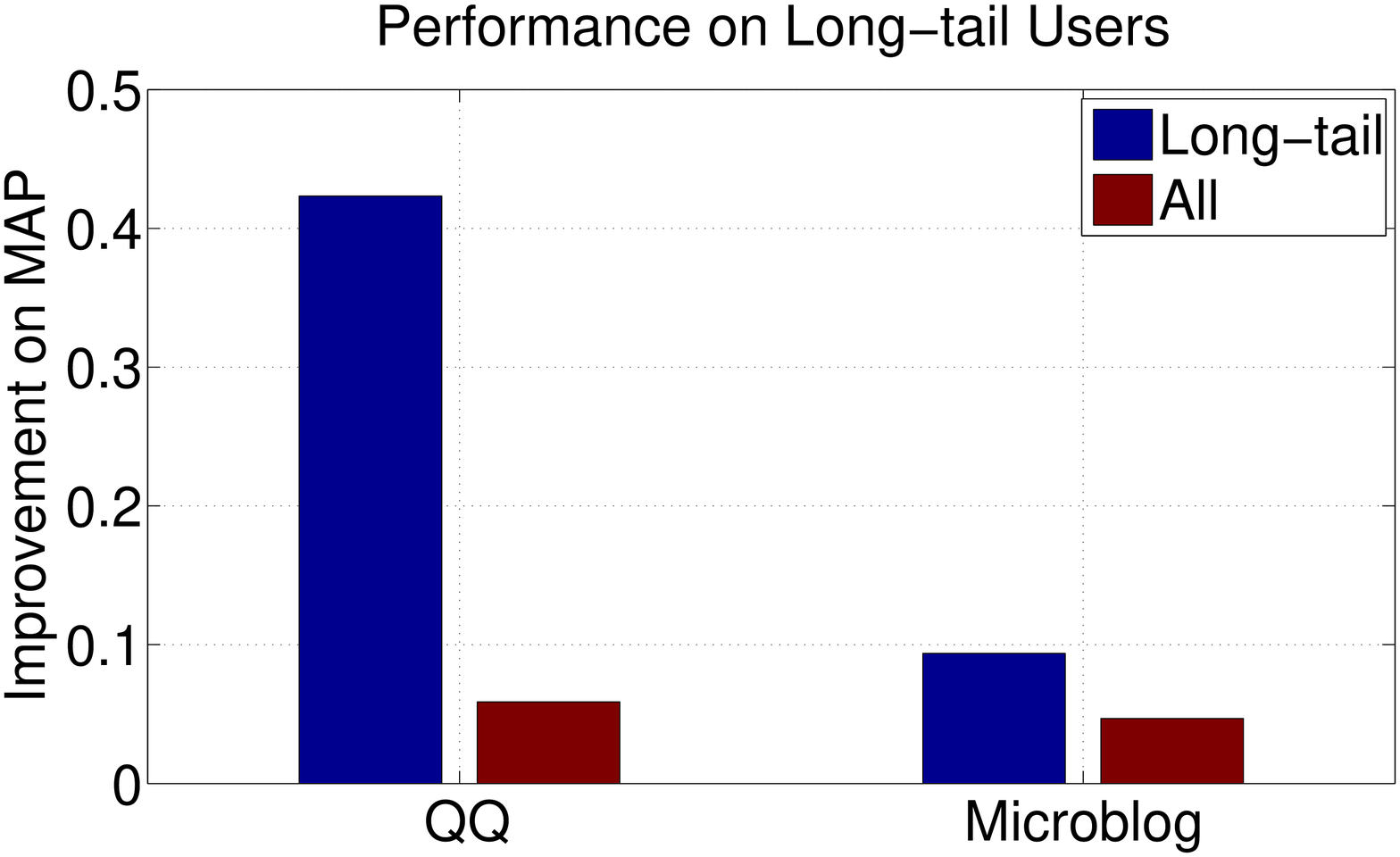}}}
\subfigure[Douban]{\scalebox{0.5}{\includegraphics[width=\columnwidth]{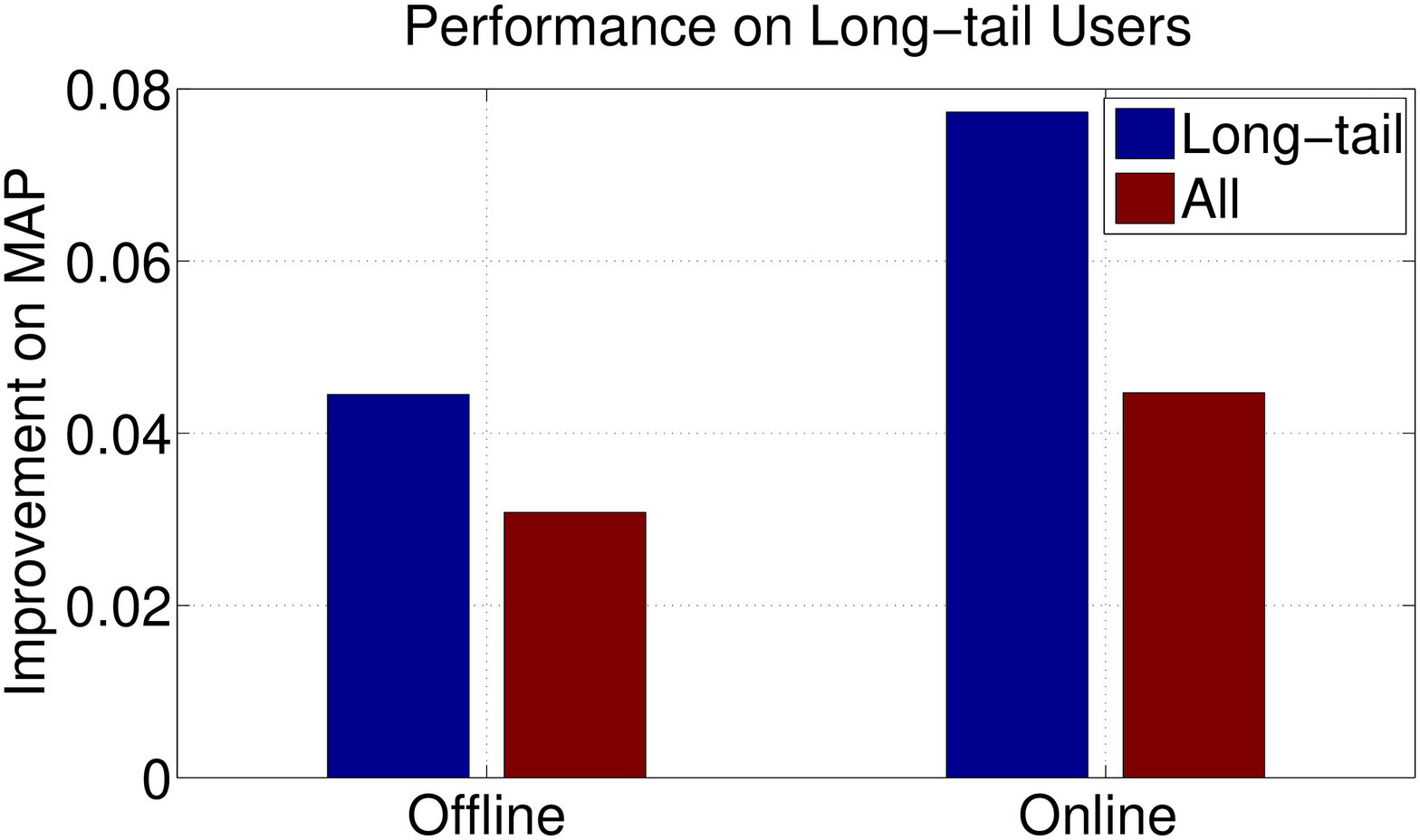}}}
\subfigure[Epinion]{\scalebox{0.5}{\includegraphics[width=\columnwidth]{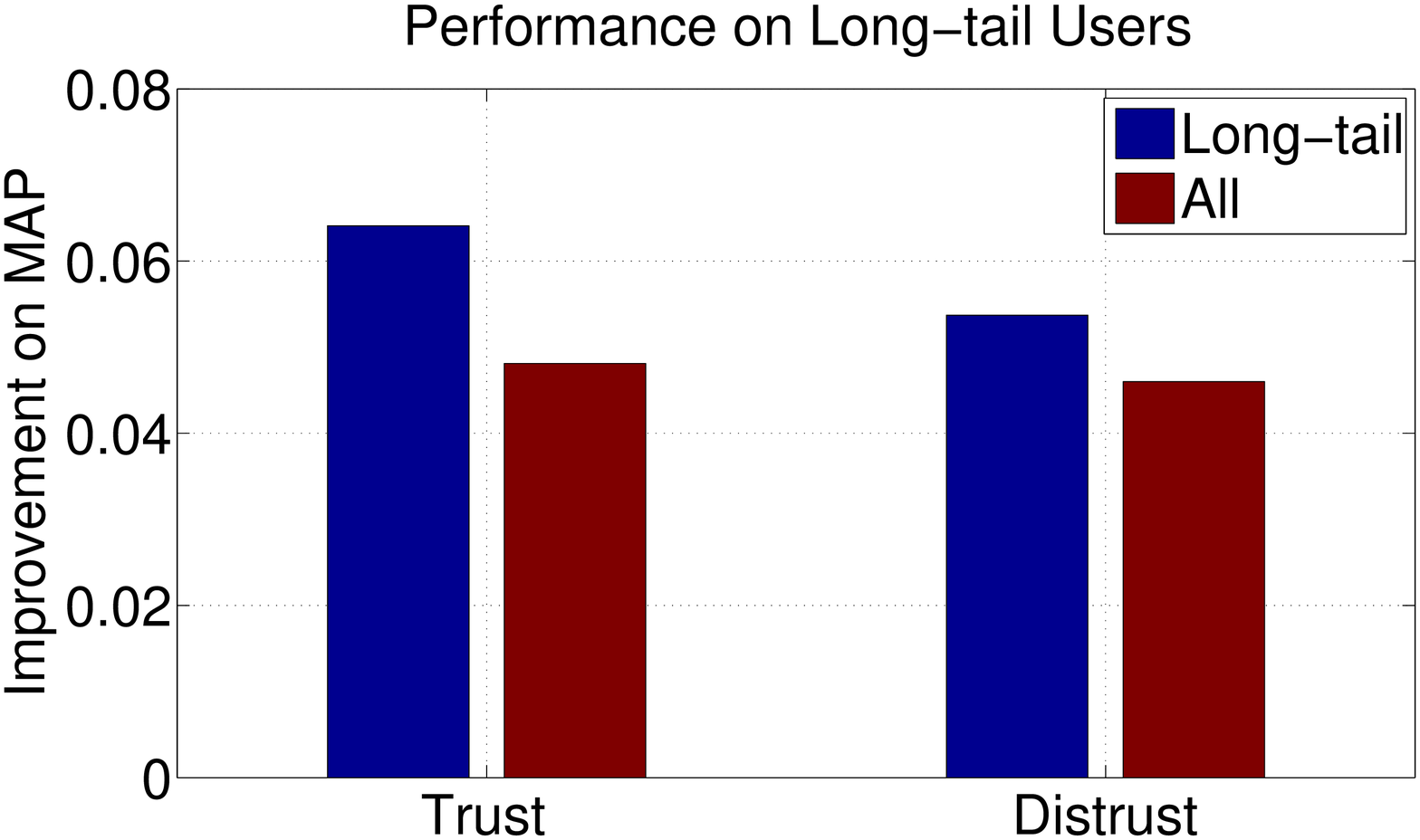}}}
}
\mbox{
\subfigure[Facebook]{\scalebox{0.5}{\includegraphics[width=\columnwidth]{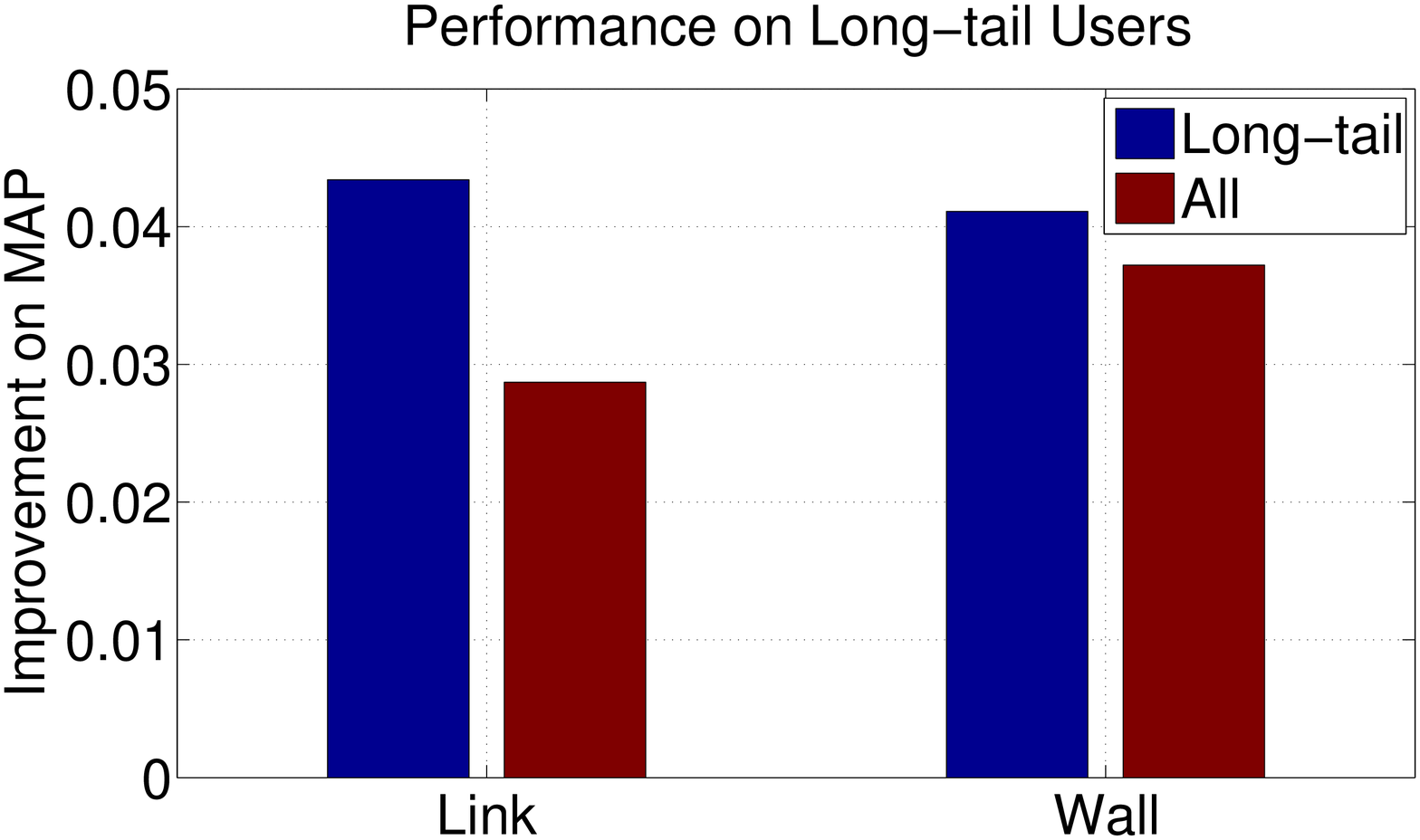}}}
\subfigure[Renren]{\scalebox{0.5}{\includegraphics[width=\columnwidth]{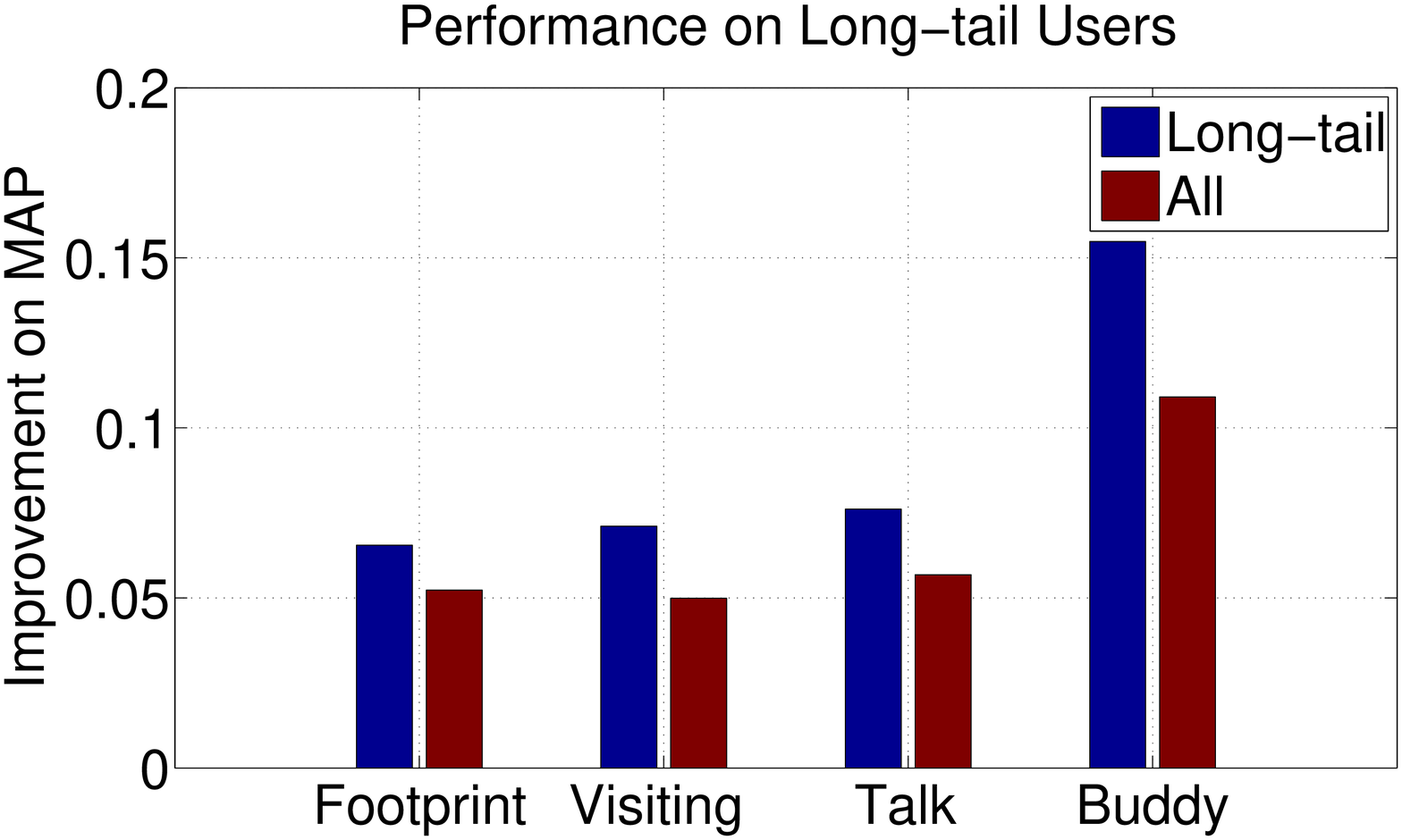}}}
\subfigure[Twitter]{\scalebox{0.5}{\includegraphics[width=\columnwidth]{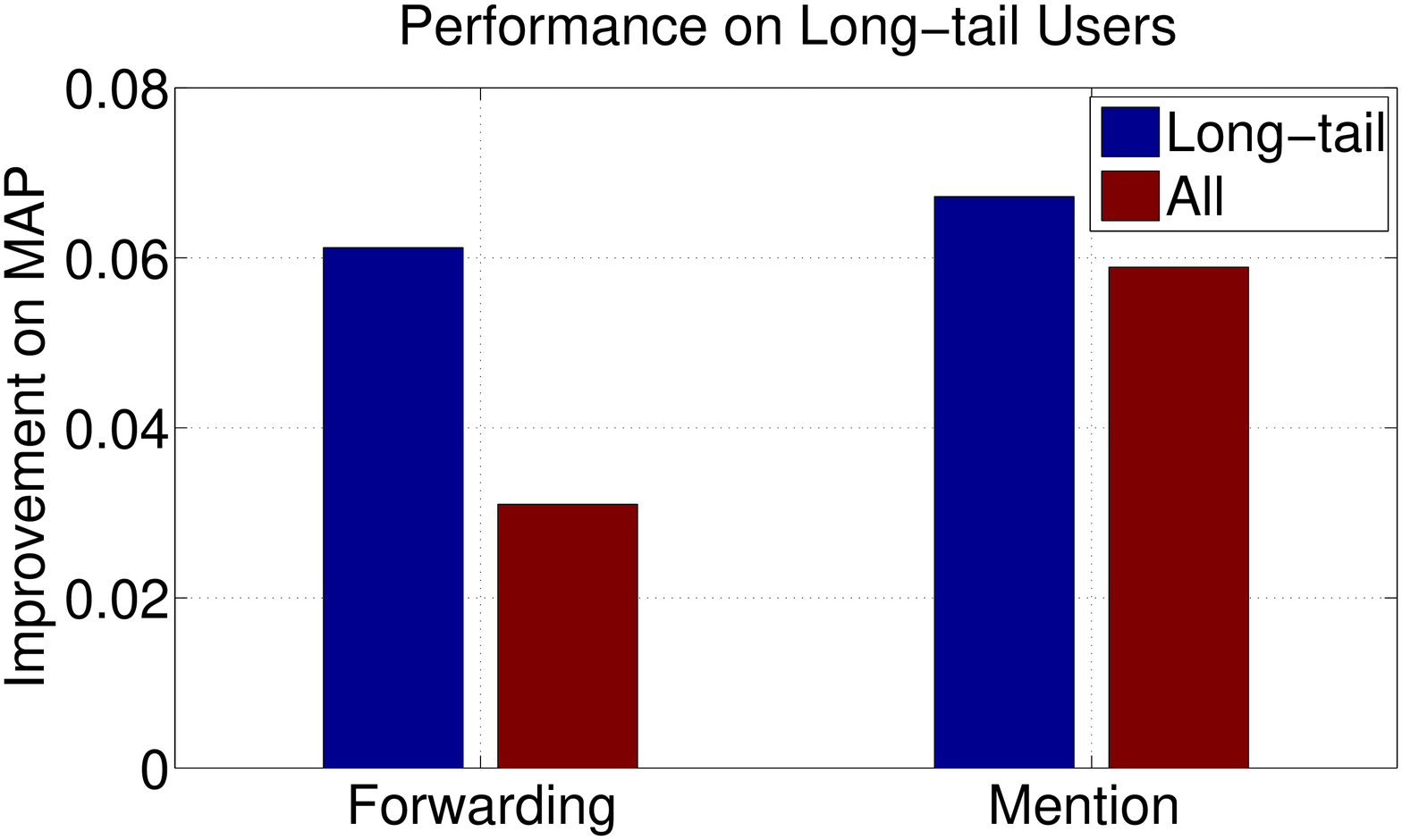}}}
}
\mbox{
\subfigure[Weibo]{\scalebox{0.5}{\includegraphics[width=\columnwidth]{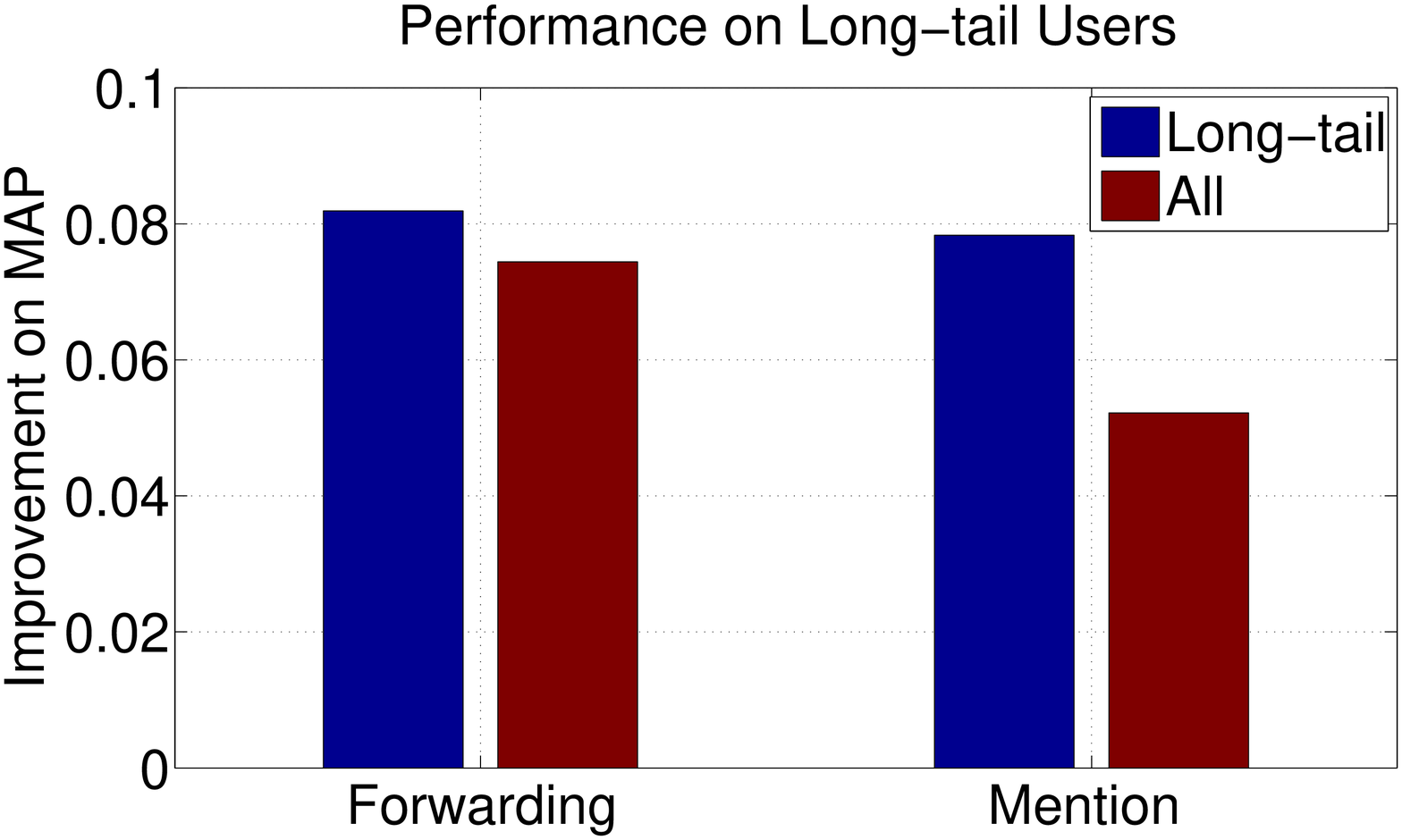}}}
\subfigure[Github]{\scalebox{0.5}{\includegraphics[width=\columnwidth]{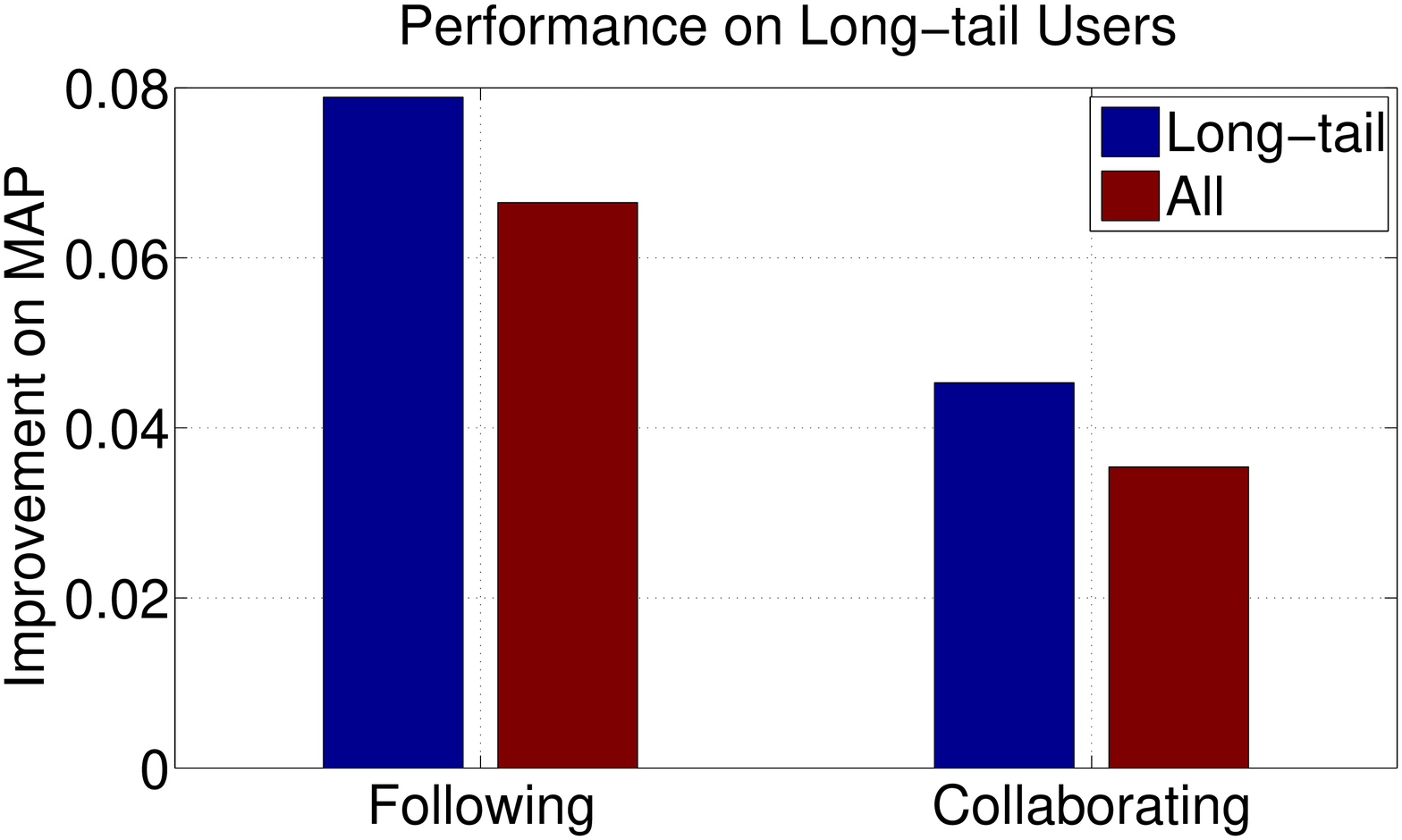}}}
\subfigure[StackOverflow]{\scalebox{0.5}{\includegraphics[width=\columnwidth]{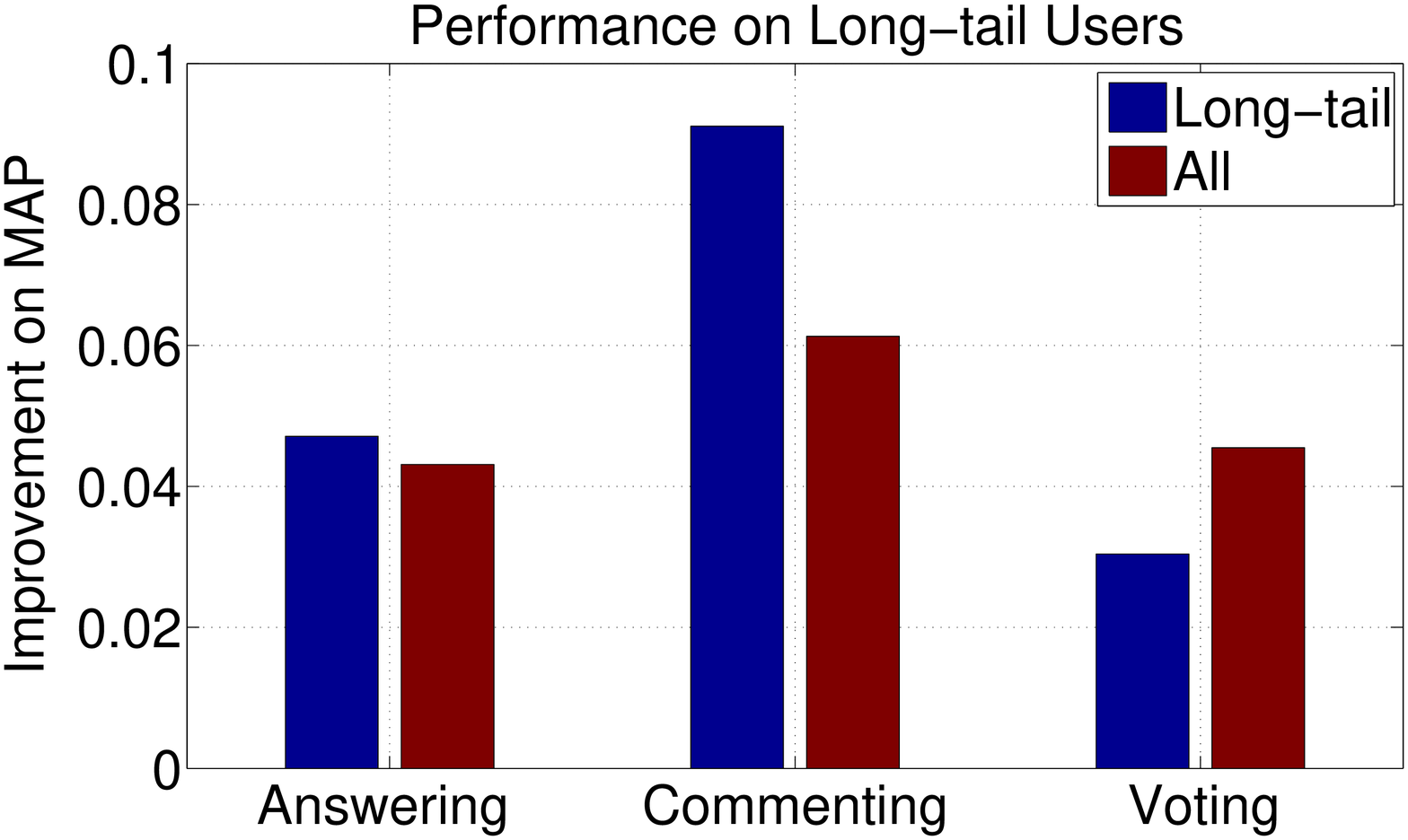}}}
}
\vskip -0.15in
\caption{\small Performances on Long-tail Users} \label{fig:longtail}
\vskip -0.14in
\end{small}
\end{figure*}

\subsection{Performance Comparisons}
We compare ComFP with other baselines on friendship prediction, and the results are summarized in Table~\ref{tab:performance}. It is evident that ComFP achieves better performance on most datasets, except the Microblog network in Tencent collection. Due to the ``sparsity'' problem, MMSB in single networks fails to capture the true users' distribution over communities under a uniform prior. For example, in Tencent collection, it achieves MAP just no more than 0.49 on QQ-IM network and no more than 0.21 on Microblog network. By considering the information from both networks, MMSB gets better performance which improves the MAP in sparser networks, such as online network of Douban collection and Link network of Facebook collection. However, it assigns a uniform prior to all single networks and thus does not tackle with network differences. This harms the performance on many other networks, such as Microblog network in Tencent collection and Offline network in Douban collection, which are comparatively denser. In addition, for those networks that are totally different, networks in Epinion collection, MMSB-C performs significantly worse than MMSB. Thus, from another side, these results support the necessity to consider network differences. Nonetheless, ComFP outperforms MMSB in all datasets. It achieves up to 0.06 higher in MAP than the baseline methods on QQ-IM network in Tencent collection. Furthermore, the improvements of ComFP on sparser networks, such as QQ-IM in Tencent collection and online network in Douban collection, are higher than other two networks. That means ComFP can bring more benefits to sparser networks. The better performance of ComFP over MMSB-C can be ascribed to the hybrid prior. As analyzed, this prior considers the user-dependent and network-specific knowledge at the same time to get benefits from auxiliary knowledge, while avoiding the harm of network differences. In addition, this prior is personalized for each user, and then makes each user choose the regularization strength from other networks adaptively. More importantly, ComFP performs better than both TF and MRLP which deal with multi-relational networks. The reason is that these two methods treat each network uniformly and do not tackle with network differences. For example, in the Epinion dataset, where the link information in two individual networks can be very different, ComFP outperforms them significantly.

{\bf Long-tail Users} To examine whether knowledge in other networks can help solve the data sparsity problem in the current network, we evaluate and illustrate the results on the long-tail users, the number of whose neighbors is smaller than $10$ and suffer from the harm of ``sparsity'' most. As shown in Figure~\ref{fig:longtail}, ComFP improves MMSB on these users more than the average level. One reason is that, for those users who have large amount of friends, MMSB has gained enough knowledge to infer their community memberships, and thus their friendships can be predicted correctly. For the long-tail users, ComFP can exploit knowledge from other networks to enhance the prediction. This also explains the larger improvements of ComFP on sparser network, such as QQ-IM of Tencent collection. From the empirical perspective, this provides justification to the argument that ComFP can enrich the knowledge in each single network.

\begin{figure}[t]
\centering \mbox{
\subfigure[Ratio]{\scalebox{0.5}{\includegraphics[width=\columnwidth]{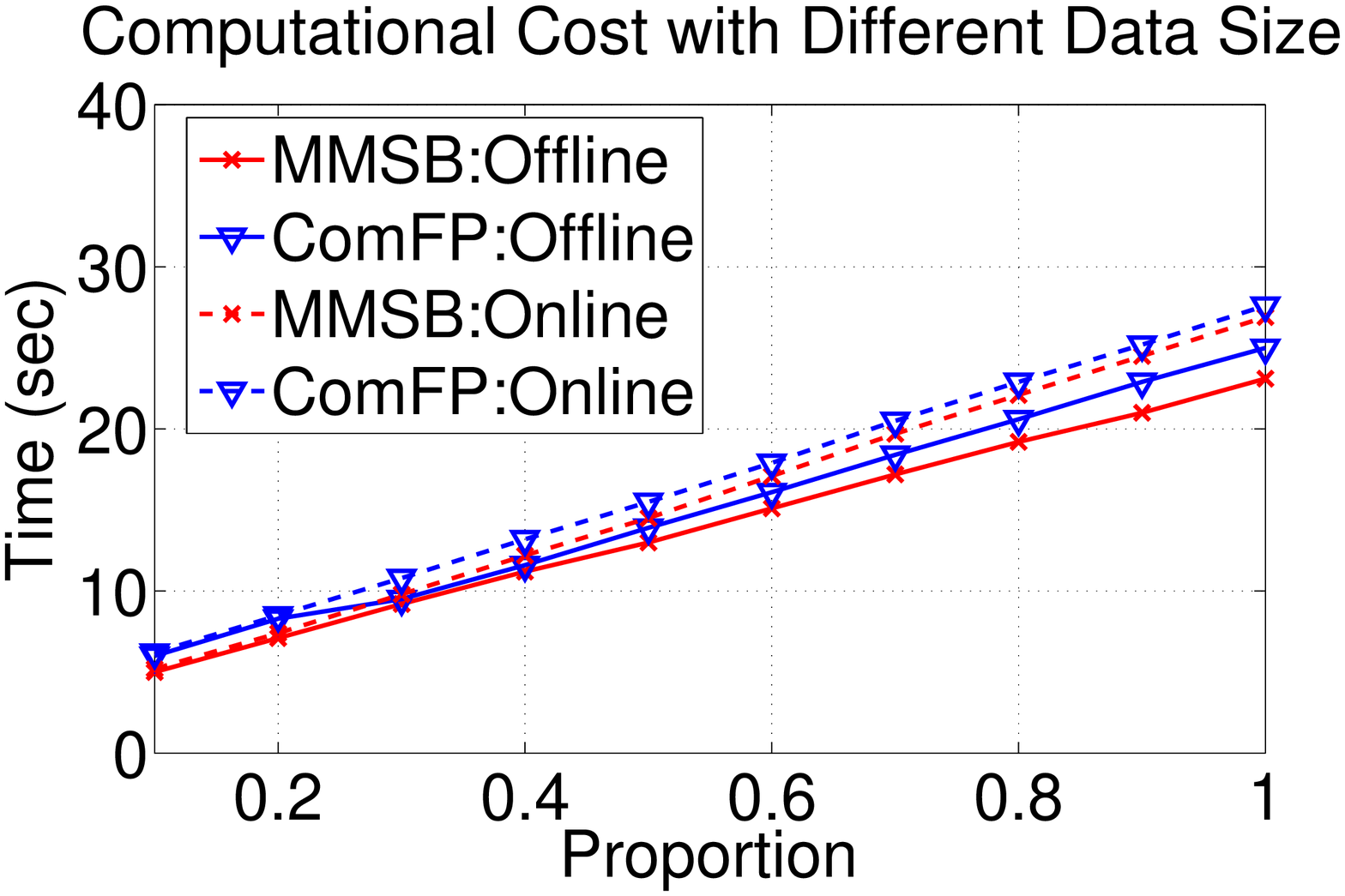}}}
\subfigure[K]{\scalebox{0.5}{\includegraphics[width=\columnwidth]{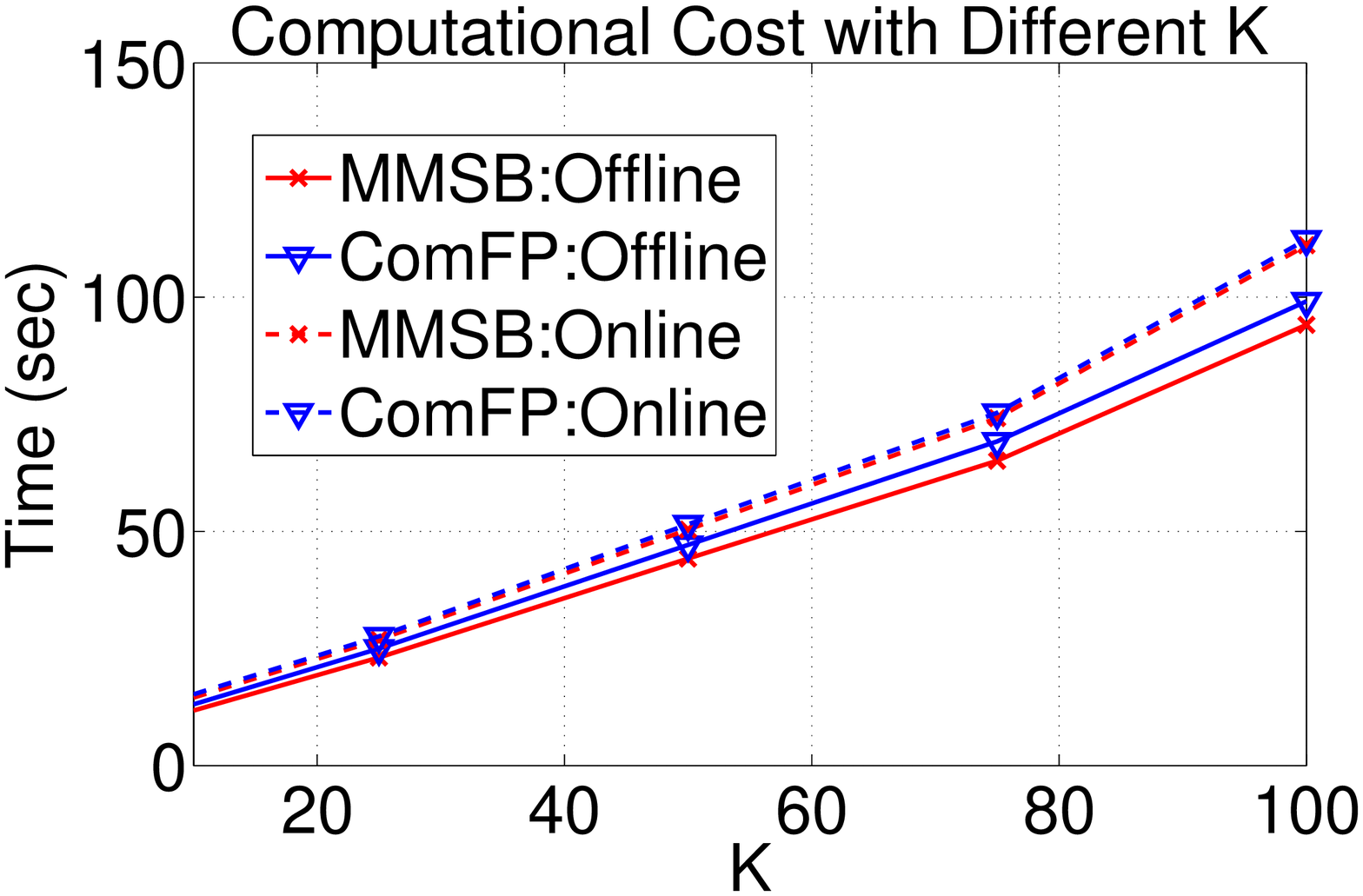}}}
}
\vskip -0.1in
\caption{\small Efficiency Analysis on Douban dataset} \label{fig:efficiency}
\vskip -0.1in
\end{figure}

\subsection{Performance Analysis}

We perform extensive experiments on Douban dataset to address the following questions: (1) Does ComFP capture network differences? (2) How do the model parameters, the corresponding ratio between two networks, and the level of data sparsity affect ComFP's performances? (3) What is the time cost of ComFP?

{\bf Network Adaptation} We illustrate the network differences captured by ComFP in Figure~\ref{fig:parameter}(a) which represents the $5\times{5}$ network-specific prior matrix $\lambda$ in each network, where the grayscale indicates the values of elements in the matrices (after normalization). Each row can be considered as a community distribution over one user feature. Obviously, communities' distributions are different between two networks. That implies ComFP works well in capturing the network differences. It also explains why C-MMSB fails to achieve good results because C-MMSB assigns a uniform prior over two networks. This result is also consistent with the example networks in Figure~\ref{fig:diff}(c) and (d), where network structures in online and offline networks can be very different, where community sizes are different and the tie strengths between different communities are also different across networks.

{\bf Parameter Analysis} We analyze the effectiveness of two parameters: one is the number of user features $T$ in the hybrid prior and the other is the number of communities $K$ in each single MMSB, on the Douban dataset. The former one indicates the complexity of user-network relations, and the latter one reflects the complexity of users' representation in each single network. We fix $K=25$ and varies $T$ from 5 to 100. The results on $T$ are illustrated in Figure~\ref{fig:parameter}(b). We observe that ComFP is not sensitive to $T$. We fix $T=25$ and let $K$ increase from 5 to 100. The results on $K$ are shown in Figure~\ref{fig:parameter}(c). Different from $T$, the performance of ComFP goes up first and then drops down when $K$ keeps increasing. That means $K$ should be tuned to avoid overfitting, such as transferred cross-validation~\cite{Zhong:2010:CVF:1889788.1889824}. In addition, we test the effectiveness of the corresponding ratio between two networks or the number of users who exist in two networks at the same time. Figure~\ref{fig:parameter}(d) presents the results. We see that, ComFP's performance becomes worse if fewer corresponds are provided across networks. However, a more important observation is that it still outperforms each MMSB in single networks consistently as long as there are enough corresponding users between two networks. Figure~\ref{fig:parameter}(e) shows the improvement of ComFP against MMSB in different levels of sparsity, where we observe that when the data become sparser in the individual network, the improvement of ComFP is more significant.

{\bf Efficiency Analysis} As analyzed in the end of Section 2.3, the computational time of ComFP increases linearly with the number of interactions between users and the number of communities $K$. We evaluate these empirically as shown in Figure~\ref{fig:efficiency}. Figure~\ref{fig:efficiency}(a) illustrates the computational time of MMSB-C and ComFP on the Douban dataset, with different ratio of links. We observe that the computational time increases linearly with larger data size. For our experimental setting, every round of inference takes about 25 seconds in our computer, of which memory is 16G and CPU is 3.2Gz. ComFP not only has better prediction performance as shown in Table~\ref{tab:performance}, but also has similar time cost with MMSB-C. Figure~\ref{fig:efficiency}(b) shows the computational time with different number of communities. Clearly, the time cost of ComFP increases linearly with $K$ as well. This suggests that ComFP can scale-up to handle large-scale datasets.

\begin{table}[t]
\caption{Related Works}
\begin{center}
\begin{footnotesize}
\begin{tabular}{|c|c|c|}
  \hline
                & Machine Learning     & Social Networks   \\
  \hline
  Single Domain & Classification, etc. & Traditional SNA   \\
  \hline
  Cross-domain  & Transfer Learning    & \emph{This work }        \\
  \hline
\end{tabular}
\label{tab:prob}
\end{footnotesize}
\end{center}
\vskip -0.15in
\end{table}

\section{Related Works}
Generally, the proposed approach in this paper can be considered as a cross-domain extension of mixed membership models and an application of transfer learning on social network analysis. The relationship between the proposed work and other previous research can be found in Table~\ref{tab:prob}. We summarized these related works in this section.

\textbf{Social Network Analysis} Recently, social network analysis has drawn lots of research interests, ranging from link prediction~\cite{Liben-Nowell:2003:LPP:956863.956972}, predicting how likely an unobserved edge exists between nodes; community detection~\cite{Leskovec:2008:SPC:1367497.1367591}, identifying communities of interest and studying their behaviors over time in social networks; to social influence~\cite{Xiang:2010:MRS:1772690.1772790}, studying how the perceived relationships with other users influence the behavior of a given user. Among them, link prediction is an active research field. Existing works predict links using pre-defined criteria, such as \emph{Katz Index}~\cite{graph:katz} and SimRank~\cite{Jeh:2002:SMS:775047.775126}, latent space models~\cite{NIPS2009_0960} and random walk~\cite{Backstrom:2011:SRW:1935826.1935914}, mixed membership models~\cite{Airoldi:2008:MMS:1390681.1442798}, etc. However, these works focus on a single network which may suffer from data sparsity problem. Several works have been proposed to handle social networks with multiple relations, such as tensor factorization~\cite{DBLP:conf/cidm/GaoDG11} and triangle patterns counting~\cite{Davis:2011:MLP:2055438.2055676}. However, these works treat each type of relationship as equally important and do not consider network differences. That may bring unnecessary auxiliary regularization to the current network.

\textbf{Transfer Learning} Transfer Learning or Domain Adaptation solves the lack of supervision problem in target applications by ``borrowing'' supervised knowledge from related problems~\cite{PanY08TLreport,Daume:2006:DAS:1622559.1622562}. It was first applied on classification, where two representative techniques are instance weighting~\cite{1273521} which filters those useless source domain data, and feature mapping~\cite{1557130} which transfers knowledge across domains through dimension reduction. Recently, transfer learning has been applied on many applications, such as metric learning~\cite{NIPS2010_0510}, sentiment classification~\cite{Daume:2006:DAS:1622559.1622562}, etc. However, they are designed for classification problems and cannot be applied on relational data. Cross-domain collaborative filtering is also introduced~\cite{Low:2011:MDU:2020408.2020434,icml2010_022}, where hierarchical Bayesian models are proposed to solve multiple domain user personalization. Knowledge in multiple social networks is exploited to help predict users' behaviors~\cite{DBLP:conf/aaai/PanAP11,Zhong:2012:CAT:2339530.2339641} recently. However, these approaches are applied on user-item interaction networks instead of social networks. In addition, traditional transfer learning focuses on borrowing knowledge. But in this paper, knowledge in each single network is incomplete, thus we need to consolidate pieces of knowledge from multiple networks instead of simply borrowing.

\textbf{Mixed Membership Models} Recently, mixed membership models have been demonstrated to be effective to model relational data, such as LDA~\cite{Blei:2003:LDA:944919.944937} and MMSB~\cite{Airoldi:2008:MMS:1390681.1442798}. The main idea is to represent each entity as a mixed membership vector over communities or topics. MMSB~\cite{Airoldi:2008:MMS:1390681.1442798} is one of the basic models. It aims to identify (i) the mixed membership mapping of users to a fixed number of communities, $K$, and (ii) the pairwise relations among the community. Then, MMSB draws links randomly between users according to the probability determined by the mixed membership and the community relations. Recently, MMSB has been extended from different aspects. For example, a hierarchical extension is proposed in~\cite{ICML2012Il_771} to utilize user features; dynamical factors are introduced in~\cite{Fu:2009:DMM:1553374.1553416, NIPS2010_0318, ChoSG11} to model temporal information; and nonparametric/infinite modeling is proposed in~\cite{Xu06}, in order to release the constraints on the number of communities. However, these approaches focus on single networks. When the data are sparse, they fail to model the mixed membership correctly due to the lack of knowledge.

\section{Conclusion and Future Works}

In this paper, we solved the friendship prediction problem across multiple nested networks, where users in different networks overlap. Each individual network is sparse and has different properties. To utilize the shared knowledge and avoid the harm due to network differences, we proposed a hierarchical Bayesian framework by introducing adaptive and hybrid priors. Unlike prior works, the proposed method ComFP considers the network-specific and user-dependent knowledge together to generate users' membership of communities. This is formulated into a hybrid prior that balances auxiliary knowledge and network differences. In addition, ComFP models users' differences adaptively with personalized priors. The proposed algorithm is flexible in that it can be extended to any number of networks and the priors can be automatically adjusted with respect to the network differences, e.g., different densities. Empirical studies were conducted on eight large-scale and real-world datasets, both of which are composed by two or more different networks, where ComFP improves previous algorithms without taking network differences and auxiliary knowledge into account by up to 0.11 in MAP.

\textbf{Scalability Discussion} While the current paper does not focus on scalability issues, our algorithms are designed with scalability in mind. For example, we can implement the proposed algorithm on Map/Reduce technologies~\cite{Dean:2008:MSD:1327452.1327492}. Following the inference framework in Algorithm~\ref{algo:MMSB} and the strategies suggested in~\cite{Smola:2010:APT:1920841.1920931}, we can design a Map and a Reduce operator in each iteration, where the Map operator computes the values of latent variables and gradients on partial data in each single machine, and the Reduce operator combines the pieces of results from each machine to update the model and users' community assignments.

\textbf{Future Works} In the future, we consider exploit users' behaviors together with relation interactions in different networks for building more accurate models. We also aim to design a model that can work more effectively under partial user correspondence among networks. Finally, we plan to apply a full Bayesian model in order to avoid overfitting on the learning of hierarchical priors.

\setlength{\bibsep}{1.0pt}
{
\small
\bibliographystyle{abbrv}
\bibliography{ref}
}

\end{document}